\documentclass[11pt]{article}

\usepackage{amsfonts, amssymb}
\usepackage{amsmath}
\usepackage{verbatim}
\usepackage{indentfirst}
\usepackage{color}

\begin{document}

\date{}

\title{Classification of $(n+2)$-dimensional $n$-Lie
Algebras
\thanks{ \small { Project partially supported by NSF(10871192) of
China, NSF(A2010000194) of Hebei Province and NSF(y2004034) of Hebei
University, China. Email address: bairp1@yahoo.com.cn;
wangxiaoling2005810805@sina.com; yzz@maths.uq.edu.au }}}
\author{\small Rui-pu  Bai$^{1,~ 3 }$  ~Xiao-ling Wang$^2$ ~Yao-zhong
Zhang$^3$
\\ \small 1. College of Mathematics and Computer,
\\\small Key Lab. in Machine Learning and Computational
Intelligence,
\\\small
Hebei University, Baoding (071002), China
\\ \small 2. College of Yuanshi, Shijiazhuang University,
 Shijiazhuang(051100), China
 \\\small 3. School of Mathematics and Physics, The University of Queensland,
\\\small Brisbane, QLD 4072, Australia}

\maketitle

\noindent{ \bf Abstract:} We give a complete classification of
$(n+2)$-dimensional $n$-Lie algebras over an algebraically closed
field of characteristic $2$, and provide a isomorphic criterion
theorem of $(n+2)$-dimensional $n$-Lie algebras.


 \noindent{\bf 2000 MR subject
classification}: 17B05 17D99

\noindent {\bf 1. Introduction}

$N$-Lie algebras are generalizations of Lie algebra.  However the
structure of $n$-Lie algebras is very different from that of Lie
algebras due to the $n$-ary multiplication. In 1985, Filippov [1]
classified $n$-Lie algebras of dimension $n+1$ over an algebraically
closed field of characteristic zero. Ling [2] proved that for every
$n\geq 3$ all finite dimensional simple $n$-Lie algebras over an
algebraically closed field $F$ of characteristic $0$ are isomorphic
to the vector product on $F^{n+1}.$ The infinite dimensional simple
$n$-Lie algebras over fields of characteristic $p\geq 0$ known
currently are only Jacobian algebras and their quotient algebras [3,
4, 5]. Bai and collaborators [6] showed  that there exist only
$[\frac{n}{2}]+1$ classes of $(n+1)$-dimensional simple $n$-Lie
algebras over a complete field of characteristic $2$ and there does
not exist simple $(n+2)$-dimensional $n$-Lie algebras.  In this
paper we give a complete classification of $(n+2)$-dimensional
$n$-Lie algebras over an algebraically closed field of
characteristic $2$.

The organization for the rest of this paper is as follows. Section 2
  introduces some basic notions, and refines the classification of
  $(n+1)$-dimensional $n$-Lie algebras given in [6] since it is the foundation for Section 3.
Section 3 describes the isomorphic criterion theorem of
$(n+2)$-dimensional $n$-Lie algebras and gives a complete
classification of $(n+2)$-dimensional $n$-Lie algebras over an
algebraically closed field of characteristic $2$.

\noindent {\bf 2. Fundamental notions}

 A vector space $A$ over a field $F$ is an $n$-Lie algebra if there
is an $n$-ary multi-linear operation $ [\ , \cdots, \ ] $ satisfying
the following identities
$$ [x_1, \cdots, x_n]=(-1)^{\tau(\sigma)}[x_{\sigma (1)}, \cdots,
x_{\sigma(n)}], \eqno(2.1) $$
 and
 $$
  [[x_1, ~\cdots, ~x_n],~
y_2, ~\cdots, ~y_n]=\sum_{i=1}^n[x_1, ~\cdots, ~[ x_i, ~y_2,
~\cdots, ~y_n], ~\cdots, ~x_n], \eqno(2.2)
$$
where $\sigma$ runs over the symmetric group $S_n$ and the number
$\tau(\sigma)$ is equal to $0$ or $1$ depending on the parity of the
permutation $\sigma.$

If the character of $F$ is $2$, then the identity (2.1) is replaced
by
$$
[x_1, \cdots, x_i, \cdots, x_j, \cdots, x_n]=0 ~\mbox{whenever}
~x_i=x_j ~\mbox{for some}~i\neq j. \eqno(2.1)'
$$

Identity (2.2) is usually called the generalized Jacobi identity, or
simply the Jacobi identity.

A derivation of an $n$-Lie algebra $A$ is a linear map $D$ of $A$
into itself satisfying
$$
 D([x_1, ~\cdots, ~x_n])=\sum_{i=1}^n[x_1, ~\cdots, ~D(x_i),
~\cdots, ~x_n], ~ \mbox{for any} ~x_1, \cdots x_n\in A.\eqno(2.3)
$$
Let $\mbox{Der}(A)$ be the set of all derivations of $A$. Then
$\mbox{Der}(A)$ is a subalgebra of the general algebra $gl(A)$ and
is called the derivation algebra of $A$.

The map  ad$(x_1, \cdots, x_{n-1})$: $A \rightarrow A$, given by
$$\mbox{ad}(x_1, \cdots, x_{n-1})(x_n)=[x_1, \cdots, x_{n}], ~\mbox{ for } x_n\in A, $$
is referred to as a left multiplication defined by elements $x_1$,
$\cdots$, $x_{n-1} \in A$. It follows from identity (2.2), that
ad$(x_1, \cdots, x_{n-1})$ is a derivation. The set of all finite
linear combinations of left multiplications is an ideal of
$\mbox{Der}(A)$, which we denote by $\mbox{ad}(A)$. Every derivation
in $\mbox{ad}(A)$ is by definition an inner derivation.

Let $A_{1}, A_{2}, \cdots, A_{n}$ be subalgebras of an $n$-Lie
algebra $A$.
 Denote by $[A_{1}, $ $A_{2},$ $\cdots, $ $A_{n}]$ the subspace of $A$ generated by all
 vectors $[x_{1},$ $ x_{2}, $ $\cdots, x_{n}]$, where $x_{i}\in A_{i}$, for $i=1, 2, \cdots, n$.
 The subalgebra $A^1 = [A, A, \cdots, A]$ is called the derived algebra of $A$. If $A^1=0,$
 then $A$ is called an abelian $n$-Lie algebra.

An ideal $I$ of an $n$-Lie algebra $A$ is a subspace of $A$ such
that $[I,$
 $ A,$ $ \cdots,$ $ A]\subseteq I. $ If $ [I,$ $ I, $ $A,$ $ \cdots, A]=0$,
 then $I$ is referred to as an abelian ideal. If $A^1\neq 0$ and $A$ has
 no ideals except $0$ and itself, then $A$ is by definition a simple
 $n$-Lie algebra.

  An $n$-Lie algebra $A$
  is said to be decomposable if there are nonzero ideals $I_1, I_2 $ such that
$$
    A = I_1\oplus I_2 ~\mbox{(direct sum as vector spaces)},
$$
then~ $[I_1, I_2, A, \cdots, A]=0 $. Otherwise, we say that $A$ is
indecomposable.
 Clearly if $A$ is a simple $n$-Lie algebra then $A$ is indecomposable.

The subset $Z(A)=\{ x\in A~|~ [x, y_1, \cdots, y_{n-1}]=0, ~\forall
y_i\in A, ~i=1, \cdots, n-1\}$ is called the center of $A$, it is
clear that $Z(A)$ is an abelian ideal of $A$.

Suppose $A$ is an $n$-Lie algebra and $V$ is a linear space, $(V,
\rho)$ is called a representation of $A$ in $V$ if the multilinear
and skew-symmetric mapping $\rho: A^{n-1}\longrightarrow
\mbox{End}(V)$ satisfies  the following identities:
$$
[\rho(a), \rho(b)]=\sum\limits_{i=1}^{n-1}\rho(b_1, \cdots,
\mbox{ad}(a)(b_i), \cdots, b_{n-1}),\eqno (2.4)
$$
$$
\rho([a_{1}, \cdots, a_{n}], b_{2}, \cdots, b_{n-1})=
\sum\limits_{i=1}^{n}(-1)^{n-i+1}\rho(a_{1}, \cdots, \hat{a}_{i},
 \cdots, a_{n})\rho(a_i, b_2, \cdots, b_{n-1}), \eqno (2.5)
$$
where $a, b\in A^{n-1}, ~a=(a_{1}, \cdots, a_{n-1}), ~ b=(b_1,
\dots, b_{n-1})$. $V$ is also referred to as an $A$-module.  Then
every $n$-Lie algebra $A$ is an $A$-module in the regular
representation $\rho(a_1,$ $ \cdots,$ $ a_{n-1})$ $=$
$\mbox{ad}(a_1, $ $\cdots, a_{n-1})$.

Let $H$ be an abelian subalgebra of $n$-Lie algebra $A$, that is
$[H, \cdots, H]=0$. Then $H$ is by definition a Toral subalgebra of
$A$, if $A$ is a complete $H$-module, that is
$$ A=\sum\limits_{\alpha \in(H^{n-1})^{\ast}}A _{\alpha}  ~ ~\mbox{(direct sum as
vector spaces)},
$$
 where
 $$ A_{\alpha}=\{x \in
A~|~\mbox{ad}(h_1, \cdots, h_{n-1})(x)=\alpha(h_1, \cdots,
h_{n-1})(x),~\forall(h_{1}, h_{2}, \cdots, h_{n-1})\in H^{n-1}\}.$$

A Toral subalgebra $H$ is called a maximal Toral subalgebra if there
are no
 Toral subalgebras of $A$ properly containing $H.$

An $n$-Lie algebra $A$ is called nilpotent, if $A$ satisfies
$A^{r}=0$ for some $r\geq 0$, where $A^{0}= A$ and $A^r$ is defined
by induction, ~$ A^{r+1}=[A^{r}, ~A, ~\cdots, ~A]$ ~for ~$r\geq 0$.

A Cartan subalgebra $H$ of  $A$ is a nilpotent subalgebra and
satisfies $N(H)=H$, where
$$
 N(H)=\{\ x\in A \mid [x, H, \cdots, H]\subseteq H\}.
 $$

In the following, unless stated otherwise, we suppose
$F$ is an algebraically closed field of characteristic $2$, and $A$
is an $n$-Lie algebra over $F$ with $n\geq 3.$ And we omit zero
brackets when listing the multiplication table in a basis of the
$n$-Lie algebra.

Firstly we give the classification of $(n+1)$-dimensional $n$-Lie
algebras over $F$. It is a simple refining of the Theorem 2.1 in
[6].

\noindent{\bf Lemma 2.1.}  Let $A$ be an $(n+1)$-dimensional $n$-Lie
algebra over $F$ and  $e_{1}, e_{2}, \cdots, e_{n+1}$ be a basis of
$A$, then one and only one of the following possibilities holds up
to isomorphism

\noindent $(a)$~ If dim$A^{1}=0$, $A$ is abelian.

\noindent$(b)$ ~If $\dim A^{1}=1$, let $A^{1} = F e_{1}$. Then  in
the case of $A^{1}\subseteq Z(A)$,
$$
 (b_{1})  ~[e_{2},  \cdots, e_{n+1}] = e_{1}.
$$

 \noindent In the case that $A^{1}$ is not contained in $Z(A)$,
$$
 (b_{2}) ~[e_{1}, \cdots, e_{n}]=e_{1}.
$$

\noindent $(c)$ ~If $\dim A^{1}=2$, let $A^{1}= F e_{1}+ F e_{2}$.
Then

$$\begin{array}{ll}
(c_{1}) ~\left\{\begin{array}{l}
{[}e_{1}, e_{3}, \cdots, e_{n+1}] = e_{2}, \\
{[} e_{2}, \cdots, e_{n+1}] = e_{1};\\
\end{array}\right.
&
 (c_{2}) ~\left\{\begin{array}{l}
{[}e_{1}, e_{3}, \cdots, e_{n+1}] = e_{2}, \\
{[} e_{2}, \cdots, e_{n+1}] = e_{1}+\beta e_{2};
\end{array}\right.
\end{array}$$
where $~ \beta\in F,$ and $ ~\beta\neq 0.$

 \noindent $(d)$ ~If $\dim A^{1}=r \geq 3$, let $A^{1}= F e_{1}+ F e_{2}+
\cdots +Fe_{r}$. Then

\vspace{2mm} \noindent $\begin{array}{l}
 (d_{1})~ ~\left\{\begin{array}{l}
{[}\hat{e}_{1}, e_{2}, \cdots, e_{n+1}] = e_{1}, \\
\cdots \cdots \cdots \cdots \cdots \cdots \cdots\\
{[}e_{1}, \cdots, \hat{e}_{p}, \cdots, e_{n+1}] = e_{p}, \\
{[}e_{1}, \cdots, \hat{e}_{p+1}, \cdots, e_{n+1}] = e_{r}, \\
{[}e_{1}, \cdots, \hat{e}_{p+2}, \cdots, e_{n+1}] = e_{r-1}, \\
\cdots \cdots \cdots \cdots \cdots \cdots \cdots\\
{[}e_{1}, \cdots, \hat{e}_{p+q}, \cdots, e_{n+1}] = e_{p+1};
\end{array}\right.
\end{array}$
$
\begin{array}{l} (d_{2}) ~\left\{\begin{array}{l}
{[}\hat{e}_{1}, e_{2}, \cdots,e_{n+1}] = e_{1}, \\
{[} e_{1}, \hat{e}_{2}, \cdots, e_{n+1}] = e_{2}, \\
\cdots \cdots \cdots \cdots \cdots \cdots \cdots\\
{[}e_{1}, \cdots, \hat{e}_{r-1}, \cdots, e_{n+1}] = e_{r-1}, \\
{[}e_{1}, \cdots, \hat{e}_{r}, \cdots, e_{n+1}] = e_{r};
\end{array}\right.
\end{array}$

 \noindent where $q$ is even, $p+q=r$ and $0< q \leq r$. And symbol $\hat{e}_i$ means that
 $e_i$ is omitted in the bracket.

 \noindent {\bf Proof.} We only need to prove the case $(c)$ since
 the other cases are proved in [6].

 By [Theorem 2.1, 6], when $\dim A^1=2$,  the multiplication
 table of $A$ in the basis $e_1, \cdots, e_{n+1}$ is given by $(c_1)'$ and
 $(c_2)$, where

\vspace{1mm} \noindent$\begin{array}{l}  (c_1)'
~\left\{\begin{array}{l}
{[}e_{1}, e_{3}, \cdots, e_{n+1}] = \alpha e_{2}, \\
{[} e_{2},  \cdots, e_{n+1}] = e_{1};
\end{array}\right.
\end{array}~ \alpha\in F, ~\alpha\neq 0.$

\vspace{1mm} \noindent Replacing $e_2$ and $e_{n+1}$ by
$\sqrt{\alpha}e_2$ and $\frac{1}{\sqrt{\alpha}}e_{n+1}$ in $(c_1)'$
respectively, we get that $(c_1)'$ is isomorphic to
$\begin{array}{l} (c_{1}) ~\left\{\begin{array}{l}
{[}e_{1},  e_{3}, \cdots, e_{n+1}] = e_{2}, \\
{[} e_{2}, \cdots, e_{n+1}] = e_{1}.
\end{array}\right.
\end{array}$

Now we prove that $n$-Lie algebras corresponding to the case $(c_2)$
with nonzero coefficients $\beta$ and $\beta'$ are isomorphic if and
only if $\beta=\beta'$.

We take a linear transformation of the basis $e_1, \cdots, e_{n+1}$
by replacing $e_1, e_2$ and $e_{n+1}$ by  $e_1+ a e_2$,
$e_1+\frac{1}{a}e_2$ and $\frac{1}{a}e_{n+1}$ respectively, then
$(c_2)$ is isomorphic to

\vspace{1mm} \noindent$\begin{array}{l} (c_{2})'
~\left\{\begin{array}{l}
{[}e_{1}, e_{3}, \cdots, e_{n+1}] = e_{1}, \\
{[} e_{2}, \cdots, e_{n+1}] = \frac{1}{a^2}e_{2};\\
\end{array}\right.
\end{array}
$ ~ where ~$a\in F, ~a+\frac{1}{a}=\beta$.

\vspace{1mm} \noindent And the structure of $A$ is completely
determined by the action of $\mbox{ad}(e_3, \cdots, e_{n+1})$ on
$A^1$.  From $(c_2)'$, the $n$-Lie algebras related to the case
$(c_2)$ with nonzero coefficients $\beta$ and $\beta'$ are
isomorphic if and only if there exist a nonzero element $s\in F$ and
a nonsingular $(2\times 2)$  matrix $B$ such that $$\left(
\begin{array}{cc}
1& 0 \\
0& \frac{1}{a^2}
\end{array}
    \right)=s B^{-1}\left(
\begin{array}{cc}
1& 0 \\
0& \frac{1}{a_1^2}
\end{array}
    \right)B, $$
    where ~$a+\frac{1}{a}=\beta$ and $~ a_1+\frac{1}{a_1}=\beta'$. This implies that the $n$-Lie
algebras corresponding to the case $(c_2)$ with nonzero coefficients
$\beta$ and $\beta'$ are isomorphic if and only if $a=a_1$, that is
$\beta=\beta'$ (it is clear that $\beta=\beta'$ if and only if
$a=a_1$).  \hfill$\Box$

 \vspace{1mm}\noindent {\bf Lemma
2.2.}$^{[7]}$  Let $A$ be an $(n+2)$-dimensional $n$-Lie algebra
over an algebraically closed field. Then there exists a subalgebra
of $A$ with codimension $1$.

\vspace{1mm}\noindent {\bf Lemma 2.3.}$^{[6]}$  Let $A$ be an
$(n+2)$-dimensional $n$-Lie algebra over $F$ and $0< \dim A^1\leq
2$. Then there exists a nonabelian $(n+1)$-dimensional subalgebra of
$A$ containing $A^1$.

\vspace{2mm} \noindent{ \bf 3. Classification of $(n+2)$-dimensional
$n$-Lie algebras }

First, we prove the  isomorphic criterion theorem for
$(n+2)$-dimensional $n$-Lie algebras over $F$. We need some symbols
for reducing our description. Suppose  $(A, [, \cdots, ]_1)$ and $(
A, [, \cdots, ]_2 )$ are $n$-Lie algebras with two $n$-ary Lie
products $[, \cdots, ]_1$ and $[, \cdots, ]_2$ on vector space $A$
and $e_{1},$ $ e_{2},$ $ \cdots, $ $e_{n+2}$ be a basis of $A$.  Set
$$
e_{i, j}=[e_{1}, \cdots, \hat{e}_{i}, \cdots, \hat{e}_{j}, \cdots,
e_{n+2}]_1=\sum\limits_{k=1}^{n+2}b^{k}_{i, j} e_{k}, ~b^{k}_{i,
j}\in F, 1\leq i < j \leq n+2, \eqno (3.1)
$$
$$B=\left(
      \begin{array}{ccccccc}
        b^{1}_{1, 2} & b^{1}_{1, 3} & \cdots & b^{1}_{1, n+2} & b^{1}_{2, 3} & \cdots & b^{1}_{n+1, n+2} \\
        b^{2}_{1, 2} & b^{2}_{1, 3} & \cdots & b^{2}_{1, n+2} & b^{2}_{2, 3} & \cdots & b^{2}_{n+1, n+2} \\
        \vdots & \vdots & \vdots & \vdots & \vdots & \vdots & \vdots \\
        b^{n+2}_{1, 2} & b^{n+2}_{1, 3} & \cdots & b^{n+2}_{1, n+2} & b^{n+2}_{2, 3} &
        \cdots & b^{n+2}_{n+1, n+2} \\
      \end{array}
    \right),
$$
then
$$(e_{1, 2}, e_{1, 3}, \cdots, e_{1, n+2}, e_{2, 3}, \cdots, e_{2, n+2}, \cdots, e_{n+1, n+2})=(e_{1},
e_{2}, \cdots, e_{n+2})B. \eqno (3.1)'
$$
The multiplication of $(A,[, \cdots, ]_1) $ is determined by
$((n+2)\times \frac{(n+1)(n+2)}{2})$ matrix $B$. And $B$ is by
definition
  the structure  matrix of $(A, [, \cdots, ]_1)$ with respect to
a basis $e_{1},$ $ e_{2},$ $ \cdots, $ $e_{n+2}$.

Similarly denote $\bar{B}$ is the structure  matrix of $(A, [,
\cdots, ]_2)$ with respect to the basis $e_{1},$ $ e_{2},$ $ \cdots,
$ $e_{n+2}$, that is
$$\bar{e}_{ij}=[e_{1}, \cdots, \hat{e}_{i}, \cdots, \hat{e}_{j},
\cdots, e_{n+2}]_2=\sum\limits_{k=1}^{n+2}\bar{b}^{k}_{i, j} e_{k},
~\bar{b}^{k}_{i, j}\in F, 1\leq i < j \leq n+2, \eqno(3.2) $$
$$
(\bar{e}_{1, 2}, \bar{e}_{1, 3}, \cdots, \bar{e}_{1, n+2},
\bar{e}_{2, 3},\cdots, \bar{e}_{2, n+2}, \cdots, \bar{e}_{n+1,
n+2})= (e_{1},  \cdots, e_{n+2})\bar{B}. \eqno (3.2)'
$$

By the above notations we have following criterion theorem.

\vspace{1mm}\noindent {\bf Theorem 3.1.} $N$-Lie algebras $(A, [,
\cdots,]_1)$ and $(A, [, \cdots,]_2)$ with products (3.1) and (3.2)
on an $(n+2)$-dimensional linear space $A$ are isomorphic if and
only if there exists a nonsingular $((n+2)\times(n+2))$ matrix
$T=(t_{i, j})$ such that
$$
B=T'^{-1}\bar{B}T_*,\eqno(3.3)
$$ where $T'$ is the transpose matrix of $T$, $T_{*}=(T_{k, l}^{i, j})$ is an
$(\frac{(n+1)(n+2)}{2}\times \frac{(n+1)(n+2)}{2})$ matrix, and
$T^{i, j}_{k, l}\in F$ is the determinant defined by (3.5) below for
$1\leq i, j, k, l\leq n+2$.

\noindent{\bf Proof.} If $n$-Lie algebra $(A, [, \cdots, ]_1)$ is
isomorphic to $(A, [, \cdots, ]_2)$ under an isomorphism $\sigma$.
Let $e_1, \cdots, e_{n+2}$ be a basis of $A$, and structure matrices
are (3.1) and (3.2) with respect to $e_1, \cdots, e_{n+2}$
respectively, that is
$$ e_{i, j}=[e_{1}, \cdots, \hat{e}_{i},
\cdots, \hat{e}_{j}, \cdots,
e_{n+2}]_1=\sum\limits_{k=1}^{n+2}b^{k}_{i, j} e_{k},~B=(b_{i,
j}^k)_{(n+2)\times\frac{(n+2)\times (n+2)}{2}};
$$
and
$$ \bar{e}_{i, j}=[e_{1}, \cdots, \hat{e}_{i}, \cdots,
\hat{e}_{j}, \cdots,
e_{n+2}]_2=\sum\limits_{k=1}^{n+2}\bar{b}^{k}_{i, j} e_{k},
~~\bar{B}=(\bar{b}_{i, j}^k)_{(n+2)\times\frac{(n+2)\times
(n+2)}{2}}.
$$

Denote $e'_i=\sigma(e_i), ~1\leq i\leq n+2$ and the nonsingular
$((n+2)\times (n+2))$ matrix $T=(t_{ij})$ is the transition matrix
of $\sigma$ with respect to the basis $e_{1}, $ $e_{2}, $ $\cdots, $
$e_{n+2},$ that is
$$
(\sigma(e_{1}),  \cdots, \sigma(e_{n+2}))=(e'_{1},  \cdots,
e'_{n+2})=(e_{1}, e_{2}, \cdots, e_{n+2})T.\eqno(3.4)$$

Then $$e_{k, l}'=[e'_{1}, ~ \cdots,
\hat{e'}_{k},~\cdots,~\hat{e'}_{l}, ~\cdots, ~e'_{n+2}]_2$$
\vspace{1mm}
$$=[\sum\limits_{m=1}^{n+2}t_{m, 1} e_{m},~\sum\limits_{m=1}^{n+2}t_{m, 2}
e_{m},~\cdots,~\sum\limits_{m=1}^{n+2}t_{m, k-1}
e_{m},~\sum\limits_{m=1}^{n+2}t_{m, k+1} e_{m},
$$
$$
\cdots,~\sum\limits_{m=1}^{n+2}t_{m, l-1}
e_{m},~\sum\limits_{m=1}^{n+2}t_{m,
l+1}e_{m},~\cdots,~\sum\limits_{m=1}^{n+2}t_{m, n+2} e_{m}]_2 $$
$$=T_{k, l}^{1, 2}\bar{e}_{1, 2}+T_{k, l}^{1, 3}\bar{e}_{1, 3}+ \cdots+T_{k, l}^{1, n+2}\bar{e}_{1, n+2}
+T_{k, l}^{2, 3}\bar{e}_{2, 3}+ \cdots+T_{k, l}^{n+1,
n+2}\bar{e}_{n+1, n+2},$$ where
$$T_{k, l}^{i, j}= \mbox{det}\left(
  \begin{array}{cccccccccc}
t_{1, 1}  & \cdots & t_{1, k-1}& t_{1, k+1}& \cdots & t_{1, l-1}& t_{1, l+1}&\cdots & t_{1, n+2} \\
t_{2, 1}  & \cdots & t_{2, k-1}& t_{2, k+1}& \cdots & t_{2, l-1}& t_{2, l+1}& \cdots & t_{2, n+2} \\
\vdots  & \vdots & \vdots & \vdots & \vdots & \vdots & \vdots & \vdots & \vdots\\
t_{i-1, 1}  & \cdots & t_{i-1, k-1}& t_{i-1, k+1}& \cdots & t_{i-1,
l-1}& t_{i-1, l+1}&\cdots
& t_{i-1, n+2} \\
t_{i+1, 1}  & \cdots & t_{i+1, k-1}& t_{i+1, k+1}& \cdots & t_{i+1,
l-1}& t_{i+1, l+1}&\cdots
& t_{i+1, n+2} \\
\vdots  & \vdots & \vdots&\vdots & \vdots & \vdots & \vdots & \vdots& \vdots \\
t_{j-1, 1}  & \cdots & t_{j-1, k-1}& t_{j-1, k+1}& \cdots & t_{j-1,
l-1}& t_{j-1, l+1}&\cdots
& t_{j-1, n+2} \\
t_{j+1, 1}  & \cdots & t_{j+1, k-1}& t_{j+1, k+1}& \cdots & t_{j+1,
l-1}& t_{j+1, l+1}&\cdots
& t_{j+1, n+2} \\
\vdots  & \vdots & \vdots&\vdots & \vdots & \vdots & \vdots & \vdots& \vdots \\
t_{n+1, 1}  &\cdots& t_{n+1, k-1}& t_{n+1, k+1}& \cdots &
t_{n+1, l-1}& t_{n+1, l+1}&\cdots & t_{n+1, n+2} \\
t_{n+2, 1}  &\cdots& t_{n+2, k-1}& t_{n+2, k+1}& \cdots & t_{n+2,
l-1}& t_{n+2, l+1}& \cdots
& t_{n+2, n+2} \\
\end{array}
\right) \eqno(3.5)
$$ \vspace{1mm}  $1\leq i< j\leq n+2,~1\leq k\neq l\leq
n+2$. Denote
$$T_{*}=\left(
                      \begin{array}{ccccccc}
T_{1, 2}^{1, 2} & T_{1, 3}^{1, 2} & \cdots & T_{1, n+2}^{1, 2} & T_{2, 3}^{1, 2} & \cdots & T_{n+1, n+2}^{1, 2} \\
 T_{1, 2}^{1, 3} & T_{1, 3}^{1, 3} & \cdots & T_{1, n+2}^{1, 3} & T_{2, 3}^{1, 3} & \cdots & T_{n+1, n+2}^{1, 3} \\
 \vdots & \vdots & \vdots & \vdots & \vdots & \vdots & \vdots \\
 T_{1, 2}^{n, n+2} & T_{1, 3}^{n, n+2} & \cdots & T_{1, n+2}^{n, n+2} & T_{2, 3}^{n, n+2} &
  \cdots & T_{n+1, n+2}^{n, n+2} \\
T_{1, 2}^{n+1, n+2} & T_{1, 3}^{n+1, n+2} & \cdots & T_{1,
n+2}^{n+1, n+2} & T_{2, 3}^{n+1, n+2} &
\cdots & T_{n+1, n+2}^{n+1, n+2} \\
 \end{array}
 \right),\eqno(3.6)
$$
then $T_{*}$ is a $(\frac{(n+1)(n+2)}{2}\times \frac{(n+1)(n+2)}{2}
)$ matrix, and
$$(e_{1, 2}', e_{1, 3}', \cdots, e_{1, n+2}', e_{2, 3}', \cdots, e_{n+1, n+2}')=
(\bar{e}_{1, 2}, \bar{e}_{1, 3}, \cdots, \bar{e}_{1, n+2},
\bar{e}_{2, 3}, \cdots, \bar{e}_{n+1, n+2})T_{*}. \eqno (3.7)$$

From identities (3.1) and (3.2)
$$ (e_{1, 2}', e_{1, 3}',
\cdots, e_{1, n+2}', e_{2, 3}', \cdots, e_{n+1, n+2}')= (e_{1},
e_{2}, \cdots, e_{n+2})\bar{B}T_{*}.\eqno(3.8)
$$ Furthermore
$$
e_{k, l}'=[e'_{1}, ~ \cdots, \hat{e}_{k}', ~\cdots, ~\hat{e}_{l}',
~\cdots, ~e'_{n+2}]_2 =[\sigma(e_1), \cdots,
\hat{\sigma(e_{k})},~\cdots,~\hat{\sigma(e_{l})}, ~\cdots,
~\sigma(e_{n+2})]_2
$$
$$
=\sigma([e_{1}, ~ \cdots, \hat{e}_{k},~\cdots,~\hat{e}_{l}, ~\cdots,
~e_{n+2}]_1) =\sigma(e_{kl})=\sum\limits_{i=1}^{n+2}b^{i}_{kl}
\sigma(e_{i})=\sum\limits_{s=1}^{n+2}(\sum\limits_{i=1}^{n+2}b^{i}_{kl})t_{si}e_s.
$$
Thus
$$
(e_{1, 2}', e_{1, 3}', \cdots, e_{1, n+2}', e_{2, 3}', \cdots,
e_{n+1, n+2}')= (e_{1}, e_{2}, \cdots, e_{n+2})T'B.\eqno(3.9)$$

It follows (3.8) and (3.9) that
$$
T'B=\bar{B}T_*, ~\mbox{that is}~B= T'^{-1}\bar{B}T_*.
$$

 On the other hand, we take a linear transformation $\sigma$ of $A$,
 such that $\sigma(e_1, \cdots, e_{n+2})=(\sigma(e_1), \cdots, \sigma(e_{n+2}))=(e_1, \cdots, e_{n+2})T.$
Similar discussion to above, we have  $\sigma$ is an $n$-Lie
isomorphism from $(A, [ , \cdots, ]_1)$ to $(A, [ , \cdots, ]_2).$~~
\hfill$\Box$

Now we give the classification theorem of $(n+2)$-dimensional
$n$-Lie algebras over  $F$.

\vspace{2mm} \noindent{\bf Theorem 3.2.} Let $A$ be an
$(n+2)$-dimensional $n$-Lie algebra over an algebraically closed
field $F$ of characteristic $2$ with a basis $e_1, e_2, \cdots,
e_{n+2}$. Then one and only one of the following possibilities
 holds up to isomorphism

\vspace{3mm} \noindent$(a)$~ If $\dim A^{1}=0$, $A$ is abelian.

\vspace{2mm} \noindent$(b)$~ If $\dim A^{1}=1$, let $A^{1}=Fe_{1}$,

\vspace{3mm} $ (b^{1}) ~{[}e_{2}, \cdots, e_{n+1}]=e_{1}$; ~
$(b^{2})~ {[}e_{1}, \cdots, e_{n}]=e_{1}.$

\vspace{3mm} \noindent$(c)$~ When $\dim A^{1}=2$, let $A^{1}=F
e_{1}+F e_{2}$. Then we have

 \noindent $
\begin{array}{ll} (c^{1}) ~\left\{\begin{array}{l}
{[}e_{2}, \cdots, e_{n+1}]=e_{1}, \\
{[}e_{1}, e_{3}, \cdots, e_{n+1}]=e_{2};
\end{array}\right.
&~~~~~~~~~~~(c^{2}) ~\left\{\begin{array}{l}
{[}e_{2}, \cdots, e_{n+1}]=e_{1}, \\
{[}e_{1}, e_{3}, \cdots, e_{n+1}]=e_{2}, \\
{[}e_{1}, e_{4}, \cdots, e_{n+2}]=e_{1}, \\
{[}e_{2}, e_{4}, \cdots, e_{n+2}]=e_{2};
\end{array}\right.
\end{array}$

\vspace{3mm} \noindent $
\begin{array}{ll}
(c^{3}) ~\left\{\begin{array}{l}
{[}e_{1},  e_3, \cdots, e_{n+1}]=e_{2}, \\
{[}e_{2}, \cdots, e_{n+1}]=e_{1}+\alpha e_{2};
\end{array}\right.
&~~~~~~(c^{4})  ~\left\{\begin{array}{l}
{[}e_{1}, e_{3}, \cdots, e_{n+1}] = e_{2}, \\
{[}e_{2}, \cdots, e_{n+1}] =e_{1}+\alpha e_{2}, \\
{[}e_{1}, e_{4}, \cdots, e_{n+2}] = e_{1}, \\
{[}e_{2}, e_{4}, \cdots, e_{n+2}] = e_{2};
\end{array}\right.
\end{array}
$

\vspace {3mm}\noindent $
\begin{array}{ll}
 (c^{5}) ~\left\{\begin{array}{l}
{[}e_{2},  \cdots, e_{n+1}]=e_{1}, \\
{[}e_{3},  \cdots, e_{n+2}]=e_{2};
\end{array}\right.
& ~~~~~~(c^{6}) ~\left\{\begin{array}{l}
{[}e_{2},  \cdots, e_{n+1}] =e_{1}, \\
{[}e_{1}, e_{4}, \cdots, e_{n+2}] = e_{1}, \\
{[}e_{2}, e_{4}, \cdots, e_{n+2}] = e_{2};
\end{array}\right.
\end{array}
$ ~ $\alpha\in F, ~\alpha\neq 0.$

\vspace{3mm} \noindent$(d)$~ If $\dim A^{1}=3$, let $A^{1}=F e_{1}+
F e_{2}+F e_{3}$. Then we have

\noindent$
\begin{array}{ll}
(d^{1}) ~\left\{\begin{array}{l}
{[}e_{1}, e_2, \hat{e}_3, \cdots, e_{n+1}]=e_{3}, \\
{[}e_{1}, e_{3}, \cdots, e_{n+1}]=e_{2}, \\
{[}e_{2},  \cdots, e_{n+1}]=e_{1};
\end{array}\right.
&~~~~~~~~~(d^{2}) ~\left\{\begin{array}{l}
{[}e_{2}, \cdots, e_{n+1}]=e_{1}, \\
{[}e_{1}, e_{3}, \cdots, e_{n+1}]=e_{3}, \\
{[}e_{1}, e_2, \hat{e}_3, \cdots, e_{n+1}]=e_{2}, \\
{[}e_{2}, e_{4}, \cdots, e_{n+2}]= e_{2}, \\
{[}e_{3}, \cdots, e_{n+2}]=e_3+ e_{2};
\end{array}\right.
\end{array}$

\vspace{3mm} \noindent$
\begin{array}{ll}
(d^{3})  ~\left\{\begin{array}{l}
{[}e_{2}, \cdots, e_{n+1}]=e_{1}, \\
{[}e_{1}, e_{3}, \cdots, e_{n+1}]=e_{3}, \\
{[}e_{1}, e_{2}, \hat{e}_3, \cdots, e_{n+1}]=e_{2};
\end{array}\right.
&~~~~~~~~~(d^{4}) ~ \left\{\begin{array}{l}
{[}e_{2}, \cdots, e_{n+1}]=e_{1}, \\
{[}e_{2}, e_{4}, \cdots, e_{n+2}]=e_{3}, \\
{[}e_{3}, \cdots, e_{n+2}]=e_{2};
\end{array}\right.
\end{array}$

\vspace{3mm}\noindent$
\begin{array}{ll}
 (d^{5})~ \left\{\begin{array}{l}
{[}e_{2}, \cdots, e_{n+1}]=e_{1}, \\
{[}e_{2}, e_{4}, \cdots, e_{n+2}]=e_{2}, \\
{[}e_{3}, \cdots, e_{n+2}]=e_{3};
\end{array}\right.
&~~~~~~~~~~~~~~(d^{6})~ \left\{\begin{array}{l}
{[}e_{2},  \cdots, e_{n+1}]=e_{1}, \\
{[}e_{1}, e_{4}, \cdots, e_{n+2}]=e_{1}, \\
{[}e_{2}, e_{4}, \cdots, e_{n+2}]=e_{2}+\gamma e_{3}, \\
{[}e_{3}, \cdots, e_{n+2}]=e_{2};
\end{array}\right.
\end{array}
$

\vspace{3mm}\noindent $\begin{array}{ll} (d^{7})~
\left\{\begin{array}{l}
{[}e_{1}, e_{4}, \cdots, e_{n+2}]=e_{1}, \\
{[}e_{2}, e_{4}, \cdots, e_{n+2}]=e_{3}, \\
{[}e_{3}, \cdots, e_{n+2}]=\beta e_2+(1+\beta)e_{3};
\end{array}\right.
&~(d^{8})~ \left\{\begin{array}{l}
{[}e_{1}, e_{4}, \cdots, e_{n+2}]=e_{1}, \\
{[}e_{2}, e_{4}, \cdots, e_{n+2}]=e_{2}, \\
{[}e_{3}, \cdots, e_{n+2}]=e_{3}; \\
\end{array}\right.
\end{array}
$

\vspace{3mm} \noindent$\begin{array}{l} (d^{9})~
\left\{\begin{array}{l}
{[}e_{1}, e_{4}, \cdots, e_{n+2}]=e_{2}, \\
{[}e_{2}, e_{4}, \cdots, e_{n+2}]=e_{3}, \\
{[}e_{3}, \cdots, e_{n+2}]=se_1+te_2+ue_{3};
\end{array}\right.
\end{array}
$

\noindent where $\beta, \gamma, s, t, u\in F,$ and  $ \beta \gamma
s\neq 0.$ And $n$-Lie algebras corresponding to $(d^{9})$ with
coefficients $s, t, u$ and $s', t', u'$ respectively are isomorphic
if and only if there exists nonzero element $\delta\in F$ such that
$s'=\delta^3 s, t'=\delta^2 t$ and $u'=\delta u.$

\vspace{3mm} \noindent $(e)$ ~If dim$A^{1}=r\geq 4$ and $r$ is even.
Let $A^{1}=F e_{1}+F e_{2}+\cdots+F e_{r}$.  Then we have

 \noindent $
\begin{array}{l}
(e^{1}) ~\left\{\begin{array}{l}
{[}\hat{e}_1, e_{2}, \cdots, e_{n+1}] = e_{1}, \\
{[}e_{1}, \hat{e}_2, \cdots, e_{n+1}] = e_{2}, \\
\cdots \cdots \cdots \cdots \cdots \cdots \cdots\\
{[}e_{1}, \cdots, \hat{e}_{p-1}, \cdots, e_{n+1}] = e_{p-1}, \\
{[}e_{1}, \cdots, \hat{e}_{p}, \cdots, e_{n+1}] = e_{p}, \\
{[}e_{1}, \cdots, \hat{e}_{p+1}, \cdots, e_{n+1}] = e_{r}, \\
{[}e_{1}, \cdots, \hat{e}_{p+2}, \cdots, e_{n+1}] = e_{r-1}, \\
\cdots \cdots \cdots \cdots \cdots \cdots \cdots\\
{[}e_{1}, \cdots, \hat{e}_{p+q}, \cdots, e_{n+1}] = e_{p+1};
\end{array}\right.
\end{array}
$ ~ where $q$ is even, $2\leq q \leq r,$ $p+q=r$;

\vspace{3mm} \noindent $
\begin{array}{ll} (e^{2})~ \left\{\begin{array}{l}
{[} \hat{e}_1, e_{2},  \cdots, e_{n+1}] = e_{1}, \\
\cdots \cdots \cdots \cdots \cdots \cdots \cdots\\
{[}e_{1}, \cdots, \hat{e}_{i}, \cdots, e_{n+1}] = e_{i}, \\
\cdots \cdots \cdots \cdots \cdots \cdots \cdots\\
{[}e_{1}, \cdots, \hat{e}_{r}, \cdots, e_{n+1}] = e_{r};
\end{array}\right.
\end{array} ~ 1\leq i\leq r;
$

\vspace{3mm}\noindent $
\begin{array}{ll}  (e^{3}) ~ \left\{\begin{array}{l}
{[} \hat{e}_1, e_{2},  \cdots, e_{n+1}] = e_{1}, \\
{[}\hat{e}_2, e_{3}, \cdots, e_{n+2}] = e_{2}, \\
\cdots \cdots \cdots \cdots \cdots \cdots \cdots\\
{[}e_{2}, \cdots, \hat{e}_{i}, \cdots, e_{n+2}] = e_{i}, \\
\cdots \cdots \cdots \cdots \cdots \cdots \cdots\\
{[}e_{2}, \cdots, \hat{e}_{r}, \cdots, e_{n+2}] = e_{r};
\end{array}\right.
\end{array} ~ 2\leq i\leq r.
$

 \vspace{3mm}
\noindent $(\bar{e})$ If $\dim A^{1}=r\geq 5$ and $r$ is odd. Let
$A^{1}=F e_{1}+F e_{2}+\cdots+F e_{r}$. Then we have

\noindent $
\begin{array}{l}
(\bar{e}^{1})  \left\{\begin{array}{l}
{[} \hat{e}_1, e_{2},  \cdots,e_{n+1}] = e_{1}, \\
{[} e_1, \hat{e}_2, \cdots,e_{n+1}] = e_{2}, \\
\cdots \cdots \cdots \cdots \cdots \cdots \cdots\\
{[}e_{1}, \cdots, \hat{e}_{p}, \cdots, e_{n+1}] = e_{p}, \\
{[}e_{1}, \cdots, \hat{e}_{p+1}, \cdots, e_{n+1}] = e_{r}, \\
{[}e_{1}, \cdots, \hat{e}_{p+2}, \cdots, e_{n+1}] = e_{r-1}, \\
\cdots \cdots \cdots \cdots \cdots \cdots \cdots\\
{[}e_{1}, \cdots, \hat{e}_{p+q}, \cdots, e_{n+1}] = e_{p+1};
\end{array}\right.
\end{array}
$
 ~$q$ is even,  $2\leq q < r$ and $p+q=r$;

\vspace{3mm} \noindent $\begin{array}{l} (\bar{e}^{2})
\left\{\begin{array}{l}
{[} \hat{e}_1, e_{2},  \cdots, e_{n+1}] = e_{1}, \\
\cdots \cdots \cdots \cdots \cdots \cdots \cdots\\
{[}e_{1}, \cdots, \hat{e}_{i}, \cdots, e_{n+1}] = e_{i}, \\
\cdots \cdots \cdots \cdots \cdots \cdots \cdots\\
{[}e_{1}, \cdots, \hat{e}_{r}, \cdots, e_{n+1}] = e_{r};
\end{array}\right.
\end{array} ~ 1\leq i\leq r;$

\vspace{2mm} \noindent $
\begin{array}{l}
(\bar{e}^{3}) \left\{\begin{array}{l}
{[}\hat{e}_{1}, e_{2}, \cdots, e_{n+1}] = e_{1}, \\
{[}e_{1}, \hat{e}_{2}, \cdots, e_{n+1}] = e_{r}, \\
\cdots \cdots \cdots \cdots \cdots \cdots \cdots\\
{[}e_{1}, \cdots, \hat{e}_{i}, \cdots, e_{n+1}] = e_{r-i+2}, \\
\cdots \cdots \cdots \cdots \cdots \cdots \cdots\\
{[}e_{1}, \cdots, \hat{e}_{r}, \cdots, e_{n+1}] = e_{2}, \\
{[}e_{3}, \cdots, e_{n+2}] = e_{3}, \\
{[} e_{2}, e_{4}, \cdots, e_{n+2}] = e_2+e_{3};
\end{array}\right.
\end{array} ~2\leq i\leq r;
$

\vspace{2mm}\noindent $
\begin{array}{l}
(\bar{e}^{4}) ~ \left\{\begin{array}{l}
{[}\hat{e}_1, e_{2}, \cdots, e_{n+1}] = e_{1}, \\
{[}e_{1}, \hat{e}_{2}, \cdots, e_{n+1}] = e_{r}, \\
\cdots \cdots \cdots \cdots \cdots \cdots \cdots\\
{[}e_{1}, \cdots, \hat{e}_{i}, \cdots, e_{n+1}] = e_{r-i+2}, \\
\cdots \cdots \cdots \cdots \cdots \cdots \cdots\\
{[}e_{1}, \cdots, \hat{e}_{r}, \cdots, e_{n+1}] = e_{2}, \\
{[}e_{3}, \cdots, e_{n+2}] = e_{2};
\end{array}\right.
\end{array} ~2\leq i\leq r;
$

\vspace{2mm}\noindent $
\begin{array}{l}
(\bar{e}^{5}) \left\{\begin{array}{l}
{[}e_2, \cdots, e_{n+1}] = e_{1}, \\
{[}\hat{e}_2, e_{3},  \cdots, e_{n+2}] = e_{2}, \\
\cdots \cdots \cdots \cdots \cdots \cdots \cdots\\
{[}e_{2}, \cdots, \hat{e}_i, \cdots, e_{n+2}] = e_{i}, \\
\cdots \cdots \cdots \cdots \cdots \cdots \cdots\\
{[}e_{2}, \cdots, \hat{e}_{r}, \cdots, e_{n+2}] = e_{r};
\end{array}\right.
\end{array} ~2\leq i\leq r;$

\vspace{2mm}\noindent $
\begin{array}{l}
(\bar{e}^{6}) ~ \left\{\begin{array}{l}
{[} e_{2}, \cdots, e_{n+1}] = e_{1}, \\
{[}\hat{e}_2, e_{3}, \cdots, e_{n+2}] = e_{3}, \\
{[}e_{2}, \hat{e}_3, \cdots, e_{n+2}] = e_{2}, \\
\cdots \cdots \cdots \cdots \cdots \cdots \cdots\\
{[}e_{2}, \cdots, \hat{e}_{r-3}, \cdots, e_{n+2}] = e_{r-2}, \\
{[}e_{2}, \cdots, \hat{e}_{r-2}, \cdots, e_{n+2}] = e_{r-3}, \\
{[}e_{2}, \cdots, \hat{e}_{r-1}, \cdots, e_{n+2}] = e_{r}, \\
{[}e_{2}, \cdots, \hat{e}_{r}, \cdots, e_{n+2}] = e_{r-1}.
\end{array}\right.
\end{array}
$

\noindent{\bf Proof.} 1. The case $(a)$ is trivial.

2. Case $(b)$. Suppose $A^1=F e_1$. By Lemma 2.1, Lemma 2.2 and
Lemma 2.3 the multiplication of $A$ in the basis $e_1, \cdots,
e_{n+2}$ has the following  possibilities

 \noindent$\begin{array}{l} (1)~ \left\{\begin{array}{l}
{[}e_{2}, \cdots, e_{n+1}] = e_{1}, \\
{[}e_{1}, \cdots, \hat{e}_{i}, \cdots,  \hat{e}_{j}, \cdots,
e_{n+2}] = b_{i, j} e_{1},  ~ b_{i, j}\in F,  ~1\leq i \neq j \leq
n+1.
\end{array}\right.
\end{array}
$

\vspace{3mm}  \noindent$\begin{array}{l}
 (2)~ \left\{\begin{array}{l}
{[}e_{1}, \cdots, e_{n}] = e_{1}, \\
{[}e_{1},  \cdots, \hat{e}_{i}, \cdots, \hat{e}_{j}, \cdots,
e_{n+2}] = b_{i, j} e_{1},  ~b_{i, j}\in F, ~1\leq i \neq j \leq
n+1.
\end{array}\right.
\end{array}
$

Substituting $e_1={[}e_{2},$ $ \cdots,$ $ e_{n+1}]$ into the other
equations of (1) and using the Jacobi identities, we get
$$b_{i, j} e_{1}=[[\hat{e}_{1}, e_{2}, \cdots, e_{n+1}], e_{2},\cdots,
\hat{e}_{i},\cdots, \hat{e}_{j},\cdots, e_{n+2}]=0,~  2\leq i \neq j
\leq n+1.
$$
Thus $b_{i, j}=0$  for $2\leq i \neq j \leq n+1$, and (1) is reduced to

\vspace{2mm}$ (1)' ~\begin{array}{ll} \left\{\begin{array}{l}
{[}e_{2}, \cdots,e_{n+1}] = e_{1}, \\
{[}\hat{e}_{2}, e_{3}, \cdots, e_{n+2}] = b_{1, 2} e_{1},\\
\cdots \cdots \cdots \cdots \cdots \cdots \cdots\\
{[}e_{2}, \cdots, \hat{e}_{j},\cdots, e_{n+2}] = b_{1, j} e_{1}, \\
\cdots \cdots \cdots \cdots \cdots \cdots \cdots\\
{[}e_{2}, \cdots, \hat{e}_{n+1},  e_{n+2}] = b_{1, n+1} e_{1};
\end{array}\right.
\end{array} \quad  2 \leq j \leq n+1.
$

Replacing $e_{n+2}$ by $e_{n+2}+\sum\limits_{j=2}^{n+1}b_{1, j}
e_{j}$ in $(1)'$, we get $ (b^1) ~{[} e_{2}, \cdots, e_{n+1}] =
e_{1}. $

The table (2) can be reduced to

\vspace{2mm} $
\begin{array}{ll}
(2)' ~\left\{\begin{array}{l}
{[}e_{1}, \cdots, e_{n}] = e_{1}, \\
{[}e_{1}, \hat{e}_2, \hat{e}_{3}, \cdots, e_{n+2}] = b_{2, 3} e_{1},  \\
\cdots \cdots \cdots \cdots \cdots \cdots \cdots\\
{[}e_{1}, \cdots, \hat{e}_{i}, \cdots, \hat{e}_{j},\cdots,
e_{n+2}] = b_{i, j} e_{1},\\
\cdots \cdots \cdots \cdots \cdots \cdots \cdots\\
 {[}e_{1}, \cdots, \hat{e}_{n-1}, \hat{e}_{n}, e_{n+1}, e_{n+2}] = b_{n-1, n}
 e_{1};
\end{array}\right.
\end{array}~ ~2 \leq i\neq j \leq n,
$

\noindent after substituting $e_{n+2}+\sum\limits_{k=1}^{n}b_{k,
n+1} e_{k}$ for $e_{n+2}$ in the table (2). Substituting
$e_1={[}e_{1},$ $ e_2,$ $ \cdots,$ $ e_{n}]$ into the other
equations of $(2)'$ and applying the Jacobi identities, we obtain
$b_{i, j}=0$, $2\leq i \neq j \leq n $. Therefore, the case (2) is
isomorphic to $ (b^2) ~{[}e_{1}, e_{2}, \cdots, e_{n}] = e_{1}.$

It is clear that the case $(b^1)$ is not isomorphic to  the case
$(b^2)$ since the derived algebra of the case $(b^1)$ is contained
in the center of $A$.

 3. If $\dim A^{1}=2$, suppose $A^1=F e_1+F e_2$.
By Lemma 2.1, Lemma 2.2 and Lemma 2.3 the multiplication table in
the basis $e_{1}, \cdots, e_{n+2}$ has the following possibilities

 \noindent$
\begin{array}{l} (1) ~\left\{\begin{array}{l}
{[}e_{2}, \cdots, e_{n+1}] = e_{1}, \\
{[}e_{1}, e_{3}, \cdots, e_{n+1}] = e_{2}, \\
{[}\hat{e}_{1}, \hat{e}_{2}, e_3, \cdots, e_{n+2}]=b^{1}_{1, 2}
e_{1}+b^{2}_{1, 2}
e_{2},\\
\cdots \cdots \cdots \cdots \cdots \cdots \cdots\\
{[}e_{1}, \cdots, \hat{e}_{i}, \cdots, \hat{e}_{j}, \cdots,
e_{n+2}]=b^{1}_{i, j} e_{1}+b^{2}_{i, j} e_{2},\\
\cdots \cdots \cdots \cdots \cdots \cdots \cdots\\
{[}e_{1}, \cdots, \hat{e}_{n}, \hat{e}_{n+1}, e_{n+2}]=b^{1}_{n,
n+1} e_{1}+b^{2}_{n, n+1} e_{2};
\end{array}\right.
\end{array}$

\vspace{1mm} \noindent$
\begin{array}{l}
(2) ~\left\{\begin{array}{l}
{[}e_{2}, \cdots, e_{n+1}] = e_{1}+\alpha e_{2}, \\
{[}e_{1}, e_{3}, \cdots, e_{n+1}] = e_{2}, \\
{[}\hat{e}_{1}, \hat{e}_{2}, e_3, \cdots, e_{n+2}]=b^{1}_{1, 2}
e_{1}+b^{2}_{1, 2}
e_{2},\\
\cdots \cdots \cdots \cdots \cdots \cdots \cdots\\
{[}e_{1}, \cdots, \hat{e}_{i}, \cdots, \hat{e}_{j}, \cdots,
e_{n+2}]=b^{1}_{i, j} e_{1}+b^{2}_{i, j} e_{2},\\
\cdots \cdots \cdots \cdots \cdots \cdots \cdots\\
{[}e_{1}, \cdots, \hat{e}_{n}, \hat{e}_{n+1}, e_{n+2}]=b^{1}_{n,
n+1} e_{1}+b^{2}_{n, n+1} e_{2};
\end{array}\right.
\end{array}$

\vspace{1mm} \noindent$\begin{array}{l}
 (3)~ \left\{\begin{array}{l}
{[}e_{2}, \cdots,e_{n+1}] = e_{1}, \\
{[}\hat{e}_{1}, \hat{e}_{2}, e_3, \cdots, e_{n+2}]=b^{1}_{1, 2}
e_{1}+b^{2}_{1, 2}
e_{2},\\
\cdots \cdots \cdots \cdots \cdots \cdots \cdots\\
{[}e_{1}, \cdots, \hat{e}_{i}, \cdots, \hat{e}_{j}, \cdots,
e_{n+2}]=b^{1}_{i, j} e_{1}+b^{2}_{i, j} e_{2},\\
\cdots \cdots \cdots \cdots \cdots \cdots \cdots\\
{[}e_{1}, \cdots, \hat{e}_{n}, \hat{e}_{n+1}, e_{n+2}]=b^{1}_{n,
n+1} e_{1}+b^{2}_{n, n+1} e_{2};
\end{array}\right.
\end{array}$

\vspace{1mm} \noindent$\begin{array}{l}
 (4) ~\left\{\begin{array}{l}
{[}e_{1}, \cdots, e_{n}] = e_{1}, \\
{[}e_{3}, \cdots, e_{n+2}]=b^{1}_{1, 2} e_{1}+b^{2}_{1, 2}
e_{2},\\
\cdots \cdots \cdots \cdots \cdots \cdots \cdots\\
{[}e_{1}, \cdots, \hat{e}_{i}, \cdots, \hat{e}_{j}, \cdots,
e_{n+2}]=b^{1}_{i,j} e_{1}+b^{2}_{i, j} e_{2},\\
\cdots \cdots \cdots \cdots \cdots \cdots \cdots\\
{[}e_{1}, \cdots, \hat{e}_{n}, \hat{e}_{n+1}, e_{n+2}]=b^{1}_{n,
n+1} e_{1}+b^{2}_{n, n+1} e_{2};
\end{array}\right.
\end{array}$

\vspace{2mm}\noindent where  ~$ 1\leq i \neq j\leq n+1;$ $\alpha\in
F,~\alpha\neq 0,$ $~b_{i, j}^k\in F, ~k=1, 2$.

Firstly imposing the Jacobi identities on  (1) for  $ \{e_{2},$ $
\cdots,$ $ \hat{e}_{j},$ $ \cdots,$ $ e_{n+2} \}$, $j=3, \cdots,
n+1$; and for $ \{e_{1},$ $ \cdots,$ $ \hat{e}_{i},$ $ \cdots,$ $
\hat{e}_{j},$ $ \cdots, e_{n+2}\}$, ~$2\leq i$ $\neq$ $j$ $\leq
n+1$, we obtain $ b^{1}_{1, j}=b^{2}_{2, j},$ $b^{2}_{1,
j}=b^{1}_{2, j}$, $3\leq j \leq n+1$, $b^{1}_{i, j}=b^{2}_{i, j}=0$,
$3\leq i\neq j\leq n+1$. Then we get the reduced form

\vspace{1mm} \noindent$
\begin{array}{l}  (1)' ~\left\{\begin{array}{l}
{[}e_{2}, \cdots, e_{n+1}] = e_{1}, \\
{[}e_{1}, e_{3}, \cdots, e_{n+1}] = e_{2}. \\
{[}e_{1}, \hat{e}_2, \hat{e}_3,  \cdots,
e_{n+2}] = b^{2}_{1, 3} e_{1}+b^{1}_{1, 3} e_{2}, \\
\cdots \cdots \cdots \cdots \cdots \cdots \cdots\\
{[}e_{1}, \hat{e}_{2}, \cdots, \hat{e}_{k}, \cdots,
e_{n+2}] = b^{2}_{1, k} e_{1}+b^{1}_{1, k} e_{2}, \\
\cdots \cdots \cdots \cdots \cdots \cdots \cdots\\
{[}e_{1}, \hat{e}_{2},  \cdots, \hat{e}_{n+1},
e_{n+2}] = b^{2}_{1, n+1} e_{1}+b^{1}_{1, n+1} e_{2}, \\
{[}e_{3}, \cdots, e_{n+2}] = b^{1}_{1, 2} e_{1}+b^{2}_{1, 2} e_{2}, \\
\cdots \cdots \cdots \cdots \cdots \cdots \cdots\\
{[}e_{2}, \cdots, \hat{e}_{j}, \cdots,
e_{n+2}] = b^{1}_{1, j} e_{1}+b^{2}_{1, j} e_{2}, \\
\cdots \cdots \cdots \cdots \cdots \cdots \cdots\\
{[}e_{2}, \cdots, \hat{e}_{n+1}, e_{n+2}] = b^{1}_{1, n+1}
e_{1}+b^{2}_{1, n+1} e_{2};
\end{array}\right.
\end{array} 3\leq k\leq n+1,~ 2\leq  j \leq n+1.
$

 \noindent Replacing  $e_{n+2}$ by
$e_{n+2}+\sum\limits_{m=2}^{n+1} b^{1}_{1, m}e_{m}+b^{2}_{1,
2}e_{1}$ in $(1)'$, we have

 \noindent$
\begin{array}{l}
(1)'' ~\left\{\begin{array}{l}
{[}e_{2}, \cdots, e_{n+1}] = e_{1}, \\
{[}e_{1}, e_{3}, \cdots, e_{n+1}] = e_{2}, \\
{[}e_{1}, \hat{e}_{2}, \hat{e}_{3},  \cdots,
e_{n+2}] = b^{2}_{1, 3} e_{1}, \\
{[}e_{1}, \hat{e}_{2}, e_{3}, \hat{e}_{4}, \cdots,
e_{n+2}] = b^{2}_{1, 4} e_{1}, \\
\cdots \cdots \cdots \cdots \cdots \cdots \cdots\\
{[}e_{1}, \hat{e}_{2}, \cdots, \hat{e}_{k}, \cdots,
e_{n+2}] = b^{2}_{1, k} e_{1}, \\
\cdots \cdots \cdots \cdots \cdots \cdots \cdots\\
{[}e_{1}, \hat{e}_{2},  \cdots, \hat{e}_{n+1},
e_{n+2}] = b^{2}_{1, n+1} e_{1}, \\
{[}e_{2}, \hat{e}_{3}, \cdots,
e_{n+2}] = b^{2}_{1, 3} e_{2}, \\
\cdots \cdots \cdots \cdots \cdots \cdots \cdots\\
{[}e_{2}, \cdots,\hat{e}_{j}, \cdots,
e_{n+2}] = b^{2}_{1, j} e_{2}, \\
\cdots \cdots \cdots \cdots \cdots \cdots \cdots\\
{[}e_{2},  \cdots, \hat{e}_{n+1}, e_{n+2}] = b^{2}_{1, n+1} e_{2};
\end{array}\right.
\end{array}~ 3\leq k, j \leq n+1.
$

\vspace{1mm} If $b^{2}_{1, i}=0$ for $3\leq i \leq n+1$,  (1) is
isomorphic to $(c^1)~
\begin{array}{l} \left\{\begin{array}{l}
{[}e_{2}, \cdots,e_{n+1}] = e_{1}, \\
{[}e_{1}, e_{3}, \cdots,e_{n+1}] = e_{2}.
\end{array}\right.
\end{array}
$

\vspace{1mm} If there exists $i$ such that $b^{2}_{1, i}\neq 0$,
$3\leq i\leq n+1,$ we might as well suppose $b^{2}_{1, 3}\neq 0$.
Then substituting $e_{3}+\sum\limits_{m=4}^{n+1} \frac{b^{2}_{1,
m}}{b^{2}_{1, 3}}e_{m}$ for $e_{3}$ and $\frac{1}{b^{2}_{1,
3}}e_{n+2}$ for $e_{n+2}$ in $(1)''$, we get

\vspace{1mm} \noindent$
\begin{array}{l} (c^{2})  \left\{\begin{array}{l}
{[}e_{2}, \cdots, e_{n+1}]=e_{1}, \\
{[}e_{1}, e_{3}, \cdots, e_{n+1}]=e_{2}, \\
{[}e_{1}, e_{4}, \cdots, e_{n+2}]=e_{1}, \\
{[}e_{2}, e_{4}, \cdots, e_{n+2}]=e_{2}.
\end{array}\right.
\end{array} $

\vspace{1mm}Secondly  imposing the Jacobi identities on $(2)$ for
$\{e_{1},$ $e_2$,
 $\cdots,$ $ \hat{e}_{i},$ $ \cdots, $ $\hat{e}_{j}, $
$\cdots,$ $ e_{n+2}\}$, we obtain $b^{1}_{i, j}=b^{2}_{i, j}=0$ if
$i, j$
 satisfy $3\leq i,
j\leq n+1$ and $i\neq j$. Taking a linear transformation of the
basis $e_1,$ $\cdots, e_{n+2}$ by replacing  $e_{n+2}$ with
$e_{n+2}+\sum\limits_{j=3}^{n+1}b^{2}_{2, j}e_{j}+b^{2}_{1,
2}e_{1}$, we get

 \noindent$
\begin{array}{l}
(2)'~ \left\{\begin{array}{l}
{[}e_{2}, \cdots, e_{n+1}] = e_{1}+\alpha e_{2}, \\
{[}e_{1}, e_{3}, \cdots, e_{n+1}] = e_{2}, \\
{[}e_{1}, \hat{e}_{2}, \hat{e}_{3}, \cdots,
e_{n+2}] = b^{1}_{2, 3} e_{1},\\
\cdots \cdots \cdots \cdots \cdots \cdots \cdots\\
{[}e_{1}, \hat{e}_{2},  \cdots, \hat{e}_{k}, \cdots,
e_{n+2}] = b^{1}_{2, k} e_{1}, \\
\cdots \cdots \cdots \cdots \cdots \cdots \cdots\\
{[}e_{1}, \hat{e}_{2},  \cdots, \hat{e}_{n+1},
e_{n+2}] = b^{1}_{2, n+1} e_{1}, \\
{[}e_{2}, \hat{e}_{3}, e_{4}, \cdots,
e_{n+2}] = b^{1}_{2, 3} e_{2}, \\
\cdots \cdots \cdots \cdots \cdots \cdots \cdots\\
{[}e_{2}, \cdots,\hat{e}_{j}, \cdots,
e_{n+2}] = b^{1}_{2, j} e_{2},\\
\cdots \cdots \cdots \cdots \cdots \cdots \cdots\\
{[}e_{2}, \cdots, \hat{e}_{n+1}, e_{n+2}] = b^{1}_{2, n+1}e_{2}, \\
{[}e_{3}, \cdots, e_{n+2}] = b^{1}_{1, 2} e_{1};
\end{array}\right.
\end{array}~3\leq k, j\leq n+1.
$

\vspace{2mm} If $b^{1}_{1, 2}=b^{1}_{2, 3}=b^{1}_{2,
4}=\cdots=b^{1}_{2, m}=\cdots=b^{1}_{2, n+1}=0$, (2) is isomorphic
to

\vspace{2mm} \noindent $\begin{array}{l}
 (c^{3}) ~
\left\{\begin{array}{l}
{[}e_{1}, e_{3}, \cdots, e_{n+1}]=e_{2},\\
{[}e_{2}, \cdots, e_{n+1}]=e_{1}+\alpha e_{2}, ~\alpha\in F,
~\alpha\neq 0.
\end{array}\right.
\end{array} $

\vspace{2mm} If $b^{1}_{1, 2}\neq 0$, but $b^{1}_{2, 3}=b^{1}_{2,
4}=\cdots=b^{1}_{2, m}=\cdots=b^{1}_{2, n+1}=0$, substituting
$\frac{1}{b^{1}_{1, 2}}e_{n+2}$ for $e_{n+2}$, we get that (2) is
isomorphic to

\vspace{2mm} \noindent$\begin{array}{l}
 (c^{3})' ~
\left\{\begin{array}{l}
{[}e_{1}, e_{3}, \cdots, e_{n+1}]=e_{2}, \\
{[}e_{2}, \cdots, e_{n+1}]=e_{1}+\alpha e_{2},\\
{[}e_{3}, \cdots, e_{n+2}] = e_{1};
\end{array}\right.
\end{array} ~\alpha\in F,
~\alpha\neq 0.$
\\ \vspace{2mm} Substituting
$e_{n+2}+\alpha e_{1}+ e_{2}$ for $e_{n+2}$ in $(c^{3})'$, we get
$(c^{3})'$ is isomorphic to $(c^{3})$.

\vspace{1mm} If there exists  $b^{1}_{2, i}\neq 0$ for some $i\geq
3$, we might as well suppose $b^{1}_{2, 3}\neq 0$. Substituting
$e_{3}+\sum\limits_{m=4}^{n+1}\frac{b^{1}_{2, m}}{b^{1}_{2,
3}}e_{m}+ \frac{b^{1}_{1, 2}}{b^{1}_{2, 3}}e_{2}$ for $e_{3}$ and
$\frac{1}{b^{1}_{2,3}}e_{n+2}$ for $e_{n+2}$ in $ (2)'$, we get

\vspace{1mm} \noindent $\begin{array}{l}
 (c^{4})\left\{\begin{array}{l}
{[}e_{1}, e_{3}, \cdots, e_{n+1}]=e_{2}, \\
{[}e_{2}, \cdots, e_{n+1}]=e_{1}+\alpha e_{2},\\
{[}e_{1}, e_{4}, \cdots, e_{n+2}] = e_{1},\\
{[}e_{2}, e_{4}, \cdots, e_{n+2}] = e_{2};
\end{array}\right.
\end{array} ~\alpha\in F,
~\alpha\neq 0.$

 \vspace{2mm} Thirdly, we study the case(3). Imposing the Jacobi identities on
$\{e_{1},$ $\hat{e}_{2},$ $\cdots,$ $\hat{e}_{j},$ $\cdots,
e_{n+2}\}$, $3\leq j \leq n+1,$  $\{e_{1}, e_{2}, \cdots,
\hat{e}_{i}, \cdots, \hat{e}_{j}, \cdots, e_{n+2}\}$, $3\leq i \neq
j\leq n+1 $, and substituting
$e_{n+2}+\sum\limits_{j=2}^{n+1}b^{1}_{1, j}e_{j}$ for $e_{n+2}$, we
get the reduced form of (3)

 \vspace{2mm}
\noindent$\begin{array}{l} (3)' ~ \left\{\begin{array}{l}
{[}e_{2}, \cdots, e_{n+1}] = e_{1}, \\
{[}e_{1}, \hat{e}_{2}, \hat{e}_{3}, \cdots, e_{n+2}] = b^{2}_{1, 3}e_{1}, \\
\cdots \cdots \cdots \cdots \cdots \cdots \cdots\\
{[}e_{1}, \hat{e}_{2},  \cdots, \hat{e}_{k}, \cdots, e_{n+2}] = b^{2}_{1, k}e_{1}, \\
\cdots \cdots \cdots \cdots \cdots \cdots \cdots\\
{[}e_{1}, \hat{e}_{2},  \cdots, \hat{e}_{n+1}, \cdots, e_{n+2}] = b^{2}_{1, n+1}e_{1}, \\
{[}e_{3}, \cdots, e_{n+2}] = b^{2}_{1, 2} e_{2},\\
\cdots \cdots \cdots \cdots \cdots \cdots \cdots\\
{[}e_{2}, \cdots, \hat{e}_{j}, \cdots,
e_{n+2}] = b^{2}_{1, j} e_{2}, \\
\cdots \cdots \cdots \cdots \cdots \cdots \cdots\\
{[}e_{2}, \cdots, \hat{e}_{n+1}, e_{n+2}] = b^{2}_{1, n+1} e_{2};
\end{array}\right.
\end{array} 3\leq k\leq n+1, 2\leq j\leq
n+1. $

\vspace{2mm}\noindent If $b^{2}_{1, 3}=\cdots=b^{2}_{1,
m}=\cdots=b^{2}_{1, n+1}=0$, and $b^{2}_{1, 2}\neq0$, replacing
$e_{n+2}$ with $\frac{1}{b^{2}_{1, 2}}e_{n+2}$ in $(3)'$, we get

\vspace{2mm}\noindent$\begin{array}{l} (c^{5})~
\left\{\begin{array}{l}
{[}e_{2}, \cdots, e_{n+1}]=e_{1},\\
{[}e_{3}, \cdots, e_{n+2}]=e_{2}.
\end{array}\right.
\end{array}
$

\vspace{1mm}If there is $i$ for $3\leq i\leq n+1$ such that
 $b^{2}_{1, i}\neq 0,$ we might as well suppose $b_{1, 3}^2\neq 0.$
  Replacing $e_{3}$ with $e_{3}+\sum\limits_{m=2, m\neq
3}^{n+1}\frac{b^{2}_{1, m}}{b^{2}_{1, 3}}e_{m}$ in $(3)'$, (3) is
isomorphic to

\vspace{2mm}\noindent$
\begin{array}{l}
 (c^{6})~ \left\{\begin{array}{l}
{[}e_{2}, \cdots, e_{n+1}] =e_{1},\\
{[}e_{1}, e_{4}, \cdots, e_{n+2}] = e_{1},\\
{[}e_{2}, e_{4}, \cdots, e_{n+2}] = e_{2}.
\end{array}\right.
\end{array}
$

\vspace{2mm} By the arguments similar to the above cases, we have
$\dim A^1=1$ in the case of (4). This is a contradiction. Therefore
the case (4) is not realized.

Thanks to
 Lemma 2.1, $n$-Lie algebras corresponding to $(c^3)$ with coefficients $\alpha$ and $\alpha'$
 are not isomorphic
 when $\alpha\neq \alpha'$.
 The proof of
the case $(c^4)$ is also similar to that of the case $(c_2)$, by
replacing $\mbox{ad}(e_3, \cdots, e_{n+1})$ in the case of $(c_2)$
by $\mbox{ad}(H)$, where $H=Fe_3+ \cdots+Fe_{n+1}+Fe_{n+2}$ is the
maximal Toral subalgebra of $(c^4)$.

Now we prove that  cases $(c^1), \cdots,$  and $ (c^6)$ represent
non-isomorphic classes. It is evident that $n$-Lie algebras of the
cases  $(c^{1})$ and $(c^{3})$ are not isomorphic to the other cases
since $(c^{1})$ and $(c^{3})$ are decomposable. And by Lemma 2.1
$(c^{1})$ is not isomorphic to $(c^{3})$.

The case $(c^{5})$ is not isomorphic to any of the cases of $(c^2),
(c^4),$ and $(c^6)$ since $(c^5)$ has a non-trivial center.

For convenience, suppose $\lambda$ is the maximum of the dimensions
of the
 Toral subalgebras. It is not
difficult to see that $\lambda=n-1$ in the case of $(c^{2})$ and
$(c^6)$; $\lambda=n$ in the case of $(c^4)$.
 And the dimension of the maximal abelian ideals of inner derivation algebra of the case
$(c^6)$ is $n$, while the maximal dimension of abelian ideals of the
inner derivation algebra of $(c^{2})$ is $n+1$. It follows at once
that $(c^i)$ is not isomorphic to $(c^j)$ when $i\neq j$ for $i,
j=2, 4, 6.$

Summarizing, we get that $(c^i)$ is not isomorphic to $(c^j)$ if
$i\neq j$ for $1\leq i, j\leq 6.$

4. ~If $\dim A^{1}=3$, suppose $A^1=F e_1+F e_2+F e_3$. Then the
multiplication table of $A$ in the  basis
 $e_1, \cdots, e_{n+2}$ has the following
possibilities

\vspace{2mm}\noindent$\begin{array}{l} (1)~  \left\{\begin{array}{l}
{[}e_{2}, \cdots, e_{n+1}] = e_{1}, \\
{[}e_{1}, e_{3}, \cdots, e_{n+1}] = e_{2}, \\
{[}e_{1}, \hat{e}_{3}, \cdots, e_{n+1}] = e_{3}, \\
{[}e_{1}, \cdots, \hat{e}_{i}, \cdots, \hat{e}_{j}, \cdots, e_{n+2}]
=b^{1}_{i, j} e_{1}+b^{2}_{i, j} e_{2}+b^{3}_{i, j} e_{3};
\end{array}\right.
\end{array}~1\leq i\neq j\leq n+1;$

\vspace{2mm}\noindent$\begin{array}{l}
 (2)~  \left\{\begin{array}{l}
{[}e_{2}, \cdots, e_{n+1}] = e_{1}, \\
{[}e_{1}, e_{3}, \cdots, e_{n+1}] = e_{3}, \\
{[}e_{1}, e_2, \hat{e}_{3}, \cdots,e_{n+1}] = e_{2}, \\
{[}e_{1}, \cdots, \hat{e}_{i}, \cdots,\hat{e}_{j}, \cdots,e_{n+2}]
=b^{1}_{ij} e_{1}+b^{2}_{i, j} e_{2}+b^{3}_{i, j} e_{3};
\end{array}\right.
\end{array}~ 1\leq i\neq j\leq n+1;$

\vspace{2mm}\noindent$\begin{array}{l} (3)~  \left\{\begin{array}{l}
{[}e_{2}, \cdots, e_{n+1}] = e_{1}, \\
{[}e_{1}, e_{3}, \cdots, e_{n+1}] = e_{2}, \\
{[}e_{1}, \cdots, \hat{e}_{i}, \cdots, \hat{e}_{j}, \cdots, e_{n+2}]
=b^{1}_{i, j} e_{1}+b^{2}_{i, j} e_{2}+b^{3}_{i, j} e_{3};
\end{array}\right.
\end{array}~ 1\leq i\neq j\leq n+1;$

\vspace{2mm}\noindent$\begin{array}{l} (4)~  \left\{\begin{array}{l}
{[}e_{2}, \cdots, e_{n+1}] = e_{1}+\alpha e_{2}, \\
{[}e_{1}, e_{3}, \cdots, e_{n+1}] = e_{2}, \\
{[}e_{1}, \cdots, \hat{e}_{i}, \cdots, \hat{e}_{j}, \cdots, e_{n+2}]
=b^{1}_{i, j} e_{1}+b^{2}_{i, j} e_{2}+b^{3}_{i, j} e_{3};
\end{array}\right.
\end{array} ~ 1\leq i\neq j\leq n+1;$

\vspace{2mm}\noindent$\begin{array}{l} (5)~ \left\{\begin{array}{l}
{[}e_{2}, \cdots, e_{n+1}] = e_{1}, \\
{[}e_{1}, \cdots, \hat{e}_{i}, \cdots, \hat{e}_{j}, \cdots, e_{n+2}]
=b^{1}_{i, j} e_{1}+b^{2}_{i, j} e_{2}+b^{3}_{i, j} e_{3};
\end{array}\right.
\end{array}~1\leq i\neq j\leq n+1;$

\vspace{2mm}\noindent$\begin{array}{l} (6)~ \left\{\begin{array}{l}
{[}e_{1}, \cdots, e_{n}] = e_{1}, \\
{[}e_{1},  \cdots, \hat{e}_{i}, \cdots, \hat{e}_{j}, \cdots,
e_{n+2}] =b^{1}_{i, j} e_{1}+b^{2}_{i, j} e_{2}+b^{3}_{i, j} e_{3};
\end{array}\right.
\end{array}~1\leq i\neq j\leq n+1;$

\vspace{2mm}\noindent$\begin{array}{l} (7)~ \left\{\begin{array}{l}
{[}e_{3}, \cdots, e_{n+2}]
=b^{1}_{1, 2} e_{1}+b^{2}_{1, 2} e_{2}+b^{3}_{1, 2} e_{3},\\
{[}e_{2}, e_{4}, \cdots, e_{n+2}]
=b^{1}_{1, 3} e_{1}+b^{2}_{1, 3} e_{2}+b^{3}_{1, 3} e_{3},\\
{[}e_{1}, e_{4}, \cdots, e_{n+2}] =b^{1}_{2, 3} e_{1}+b^{2}_{2, 3}
e_{2}+b^{3}_{2, 3} e_{3}.
\end{array}\right.
\end{array}$

\vspace{1mm}Firstly we study the table (1). Substituting $e_1, e_2,
e_3$ into the other equations of (1) and  using the Jacobi
identities for $\{e_{1}, e_{4},\cdots, e_{n+2}\}$, $\{e_{2}, e_{4},$
$\cdots,$ $ e_{n+2}\}$, $\{e_{3},$ $\cdots,$ $e_{n+2}\}$, $\{e_{1},$
$e_2,$ $\cdots,$ $\hat{e}_{i},$ $\cdots,$ $\hat{e}_{j},$ $\cdots,$
$e_{n+2}\}$, $3\leq i\neq j \leq n+1$, $\{e_{1},$ $e_{3},$ $\cdots,$
$\hat{e}_{j},$ $\cdots,$ $e_{n+2}\}$, $4\leq j \leq n+1$, and
$\{e_{2},$ $e_3$, $\cdots,$ $\hat{e}_{j},$ $\cdots,$ $e_{n+2}\}$,
$4\leq j \leq n+1$, we get

\vspace{2mm}\noindent$
\begin{array}{l} (1)'~
\left\{\begin{array}{l}
{[}e_{2}, \cdots, e_{n+1}] = e_{1}, \\
{[}e_{1}, e_{3}, \cdots, e_{n+1}] = e_{2}, \\
{[}e_{1}, e_{2}, \hat{e}_{3}, \cdots, e_{n+1}] = e_{3}, \\
{[}e_{2}, \hat{e}_{3}, \cdots, e_{n+2}] = b^{1}_{1, 3} e_{1}+b^{1}_{2, 3} e_{2}+b^{2}_{1, 2} e_{3},  \\
{[}e_{2}, e_{3}, \hat{e}_{4}, \cdots, e_{n+2}] = b^{1}_{1, 4} e_{1}, \\
\cdots \cdots \cdots \cdots \cdots \cdots \cdots\\
{[}e_{2}, e_3, \cdots, \hat{e}_{k}, \cdots, e_{n+2}] = b^{1}_{1, k} e_{1}, \\
\cdots \cdots \cdots \cdots \cdots \cdots \cdots\\
{[}e_{2}, e_3 \cdots, \hat{e}_{n+1}, \cdots, e_{n+2}] = b^{1}_{1, n+1} e_{1}, \\
{[}e_{1}, \hat{e}_{2}, \hat{e}_{3},  \cdots, e_{n+2}] = b^{1}_{2, 3} e_{1}+b^{1}_{1, 3} e_{2}+b^{1}_{1, 2} e_{3},  \\
{[}e_{1}, \hat{e}_{2}, e_{3}, \hat{e}_{4}, \cdots, e_{n+2}] = b^{1}_{1, 4} e_{2}, \\
\cdots \cdots \cdots \cdots \cdots \cdots \cdots\\
{[}e_{1}, \hat{e}_{2},  \cdots, \hat{e}_{j}, \cdots, e_{n+2}] = b^{1}_{1, j} e_{2}, \\
\cdots \cdots \cdots \cdots \cdots \cdots \cdots\\
{[}e_{1}, \hat{e}_{2},  \cdots, \hat{e}_{n+1}, \cdots, e_{n+2}] =
b^{1}_{1, n+1}
e_{2}, \\
{[}e_{3},  \cdots, e_{n+2}] = b^{1}_{1, 2} e_{1}+b^{2}_{1, 2}
e_{2}+b^{1}_{2, 3} e_{3};
\end{array}\right.
\end{array}~4\leq k, j\leq n+1.
$

\vspace{1mm} Substituting $e_{n+2}+\sum\limits_{i=4}^{n+1}b^{1}_{1,
i} e_{i}+b^{2}_{1, 2} e_{1}+b^{1}_{1, 2} e_{2}+b^{1}_{1, 3} e_{3}$
for $e_{n+2}$ in $(1)'$, then (1) is isomorphic to

\vspace{2mm}\noindent $\begin{array}{l} (1)'' ~
\left\{\begin{array}{l}
{[}e_{2}, \cdots, e_{n+1}] = e_{1}, \\
{[}e_{1}, e_{3}, \cdots, e_{n+1}] = e_{2}, \\
{[}e_{1}, e_{2}, \hat{e}_{3}, \cdots, e_{n+1}] = e_{3},\\
{[}e_{1}, e_{4}, \cdots, e_{n+2}] = b^{1}_{2, 3}e_{1}, \\
{[}e_{2}, e_{4}, \cdots, e_{n+2}] = b^{1}_{2, 3}e_{2}, \\
{[}e_{3}, e_{4}, \cdots, e_{n+2}] = b^{1}_{2, 3}e_{3} .
\end{array}\right.
\end{array}$

\vspace{2mm}Since $b^{1}_{2, 3}e_{1}={[}e_{1}, e_{4}, \cdots,
e_{n+2}]= [{[}e_{2}, \cdots, e_{n+1}], e_{4}, \cdots, e_{n+2}]=0,$
the case (1) is isomorphic to $\begin{array}{l}
 (d^{1}) ~\left\{\begin{array}{l}
{[}e_{2}, \cdots, e_{n+1}] = e_{1}, \\
{[}e_{1}, e_{3}, \cdots, e_{n+1}] = e_{2}, \\
{[}e_{1}, e_{2}, \hat{e}_{3}, \cdots, e_{n+1}] = e_{3}.
\end{array}\right.
\end{array}$

\vspace{2mm}Secondly by similar discussions to (1), imposing the
Jacobi identities on  $(2)$ for $\{e_{1}, e_{4}, \cdots, e_{n+2}\}$,
$\{e_{2}, e_{4}, \cdots, e_{n+2}\}$, $\{e_{3},$ $ \cdots,
e_{n+2}\}$, $\{e_{1},$ $ e_2,$ $\cdots, $ $\hat{e}_{i},$ $\cdots,
\hat{e}_{j},$ $\cdots,$ $e_{n+2}\}$, $3\leq i$ $\neq j \leq n+1,$
$\{e_{1},$ $ e_{3},$ $\cdots,$ $\hat{e}_{j},$ $\cdots,$ $e_{n+2}\}$
and $\{e_{2},$ $e_3,$ $\cdots,$ $\hat{e}_{j},$ $\cdots,$ $
e_{n+2}\}$, $4\leq j \leq n+1$,  we get

\vspace{2mm}\noindent $\begin{array}{l} (2)' ~
\left\{\begin{array}{l}
{[}e_{2}, \cdots, e_{n+1}] = e_{1}, \\
{[}e_{1}, e_{3}, \cdots, e_{n+1}] = e_{3}, \\
{[}e_{1}, e_{2}, \hat{e}_{3}, \cdots, e_{n+1}] = e_{2}, \\
{[}e_{2}, \hat{e}_{3}, \cdots, e_{n+2}] = b^{1}_{1, 3} e_{1}+b^{2}_{1, 3} e_{2}+b^{3}_{1, 3} e_{3}, \\
{[}e_{2}, e_{3}, \hat{e}_{4}, \cdots, e_{n+2}] = b^{1}_{1, 4} e_{1}, \\
\cdots \cdots \cdots \cdots \cdots \cdots \cdots\\
{[}e_{2}, \cdots, \hat{e}_{k}, \cdots, e_{n+2}] = b^{1}_{1, k} e_{1}, \\
\cdots \cdots \cdots \cdots \cdots \cdots \cdots\\
{[}e_{2},  \cdots, \hat{e}_{n+1}, \cdots, e_{n+2}] = b^{1}_{1, n+1} e_{1}, \\
{[}e_{1}, \hat{e}_{2}, \hat{e}_{3},  \cdots, e_{n+2}] = b^{1}_{1, 3} e_{3}+b^{1}_{1, 2} e_{2},  \\
{[}e_{1}, \hat{e}_{2}, e_{3}, \hat{e}_{4}, \cdots, e_{n+2}] = b^{1}_{1, 4} e_{3}, \\
\cdots \cdots \cdots \cdots \cdots \cdots \cdots\\
{[}e_{1}, \hat{e}_{2}, \cdots, \hat{e}_{j}, \cdots, e_{n+2}] = b^{1}_{1, j} e_{3}, \\
\cdots \cdots \cdots \cdots \cdots \cdots \cdots\\
{[}e_{1}, \hat{e}_{2}, \cdots, \hat{e}_{n+1}, \cdots, e_{n+2}] =
b^{1}_{1, n+1}
e_{3},\\
{[}e_{3}, \cdots, e_{n+2}] = b^{1}_{1, 2} e_{1}+b^{2}_{1, 2}
e_{2}+b^{3}_{1, 2} e_{3};
\end{array}\right.
\end{array}~4\leq k, j\leq n+1.$

\noindent Substituting  $e_{n+2}+\sum\limits_{i=2}^{n+1}b^{1}_{1, i}
e_{i}+b^{3}_{1, 2} e_{1}$ for $e_{n+2}$ in $(2)'$, we obtain

\vspace{2mm}\noindent$
\begin{array}{l} (2)''
~\left\{\begin{array}{l}
{[}e_{2}, \cdots, e_{n+1}] = e_{1}, \\
{[}e_{1}, e_{3}, \cdots, e_{n+1}] = e_{3}, \\
{[}e_{1}, e_{2}, \hat{e}_{3}, \cdots, e_{n+1}] = e_{2}, \\
{[}e_{2}, e_{4}, \cdots, e_{n+2}] = b^{2}_{1, 3} e_{2}+b^{3}_{1, 3} e_{3}, \\
{[}e_{3}, \cdots, e_{n+2}] = b^{2}_{1, 2} e_{2}.
\end{array}\right.
\end{array}
$

\vspace{2mm} If $b^{2}_{1, 2}\neq 0$ and $ b^{3}_{1, 3}\neq 0$, we
have $b^{2}_{1, 3}=0$ since
$$[b^{2}_{1, 3} e_{2}+ b^{3}_{1, 3} e_{3}, e_{3},  \cdots,
e_{n+1}]=b^{2}_{1, 3} e_{1}=[[e_{2}, e_{4}, \cdots, e_{n+2}], e_{3},
\cdots, e_{n+1}]=0.$$  Replacing $e_2$ and $e_{n+2}$ by
$e_2+\frac{1}{\sqrt{\frac{b^{2}_{1, 2}}{b^{3}_{1, 3}}}}e_3$ and
$\frac{1+\sqrt{\frac{b^{2}_{1, 2}}{b^{3}_{1,
3}}}}{\sqrt{\frac{b^{2}_{1, 2}}{b^{3}_{1,
3}}}}e_1+\frac{1}{b^{2}_{1, 2}}e_{n+2}$ in $(2)''$, we have

\vspace{2mm}\noindent $\begin{array}{l} (d^2) ~
\left\{\begin{array}{l}
{[}e_{2}, \cdots, e_{n+1}]=e_{1}, \\
{[}e_{1}, e_{3}, \cdots, e_{n+1}]=e_{3}, \\
{[}e_{1}, e_{2}, \hat{e}_{3}, \cdots, e_{n+1}]=e_{2}, \\
{[}e_{2}, e_{4}, \cdots, e_{n+2}]=e_{2}, \\
{[}e_{3}, \cdots, e_{n+2}]=e_3+ e_{2}.
\end{array}\right.
\end{array}$

\vspace{1mm} If $b^{2}_{1, 2}=b^{3}_{1, 3}= 0$, it implies that
$(2)$ is isomorphic to $\begin{array}{l} (d^{3})~
\left\{\begin{array}{l}
{[}e_{2},  \cdots, e_{n+1}]=e_{1}, \\
{[}e_{1}, e_{3}, \cdots, e_{n+1}]=e_{3}, \\
{[}e_{1}, e_{2}, \hat{e}_{3}, \cdots, e_{n+1}]=e_{2}.
\end{array}\right.
\end{array}$

\vspace{2mm} If $b^{2}_{1, 2}\neq 0 $ and $ b^{3}_{1, 3}= 0$, from
$[b^{2}_{1, 3}$ $e_{2},$ $e_{3},$ $\cdots,$ $e_{n+1}]=b^{2}_{1, 3}
e_{1}=[[e_{2},$ $e_{4},$ $\cdots,$ $e_{n+2}],$ $e_{3},
 \cdots,$ $ e_{n+1}]=0,$~ we obtain  $b^{2}_{1, 3}=0$. Replacing $e_{n+2}$ by
$\frac{1}{b^{2}_{1, 2}}e_{n+2}+e_{1}$ in $(2)''$, we get that (2) is
isomorphic to $(d^{2})$.

Thirdly by imposing the Jacobi identities on $(3)$ for $\{e_{1},
\hat{e}_2,$  $\cdots,$ $ \hat{e}_{i},$ $ \cdots,$ $ e_{n+2}\}, $
$3\leq i \leq n+1,$ $\{e_{1},$ $ e_{2}, $ $\hat{e}_3,$
 $ \cdots,$ $ \hat{e}_{k},$ $ \cdots,$ $ e_{n+2}\}$, $4\leq
k \leq n+1$ and $\{e_{2}, b^{3}_{1, 2}e_{3}, \cdots, e_{n+1}\}$,
  we have

\vspace{2mm}\noindent$\begin{array}{l}  \left\{\begin{array}{l}
{[}e_{2}, \cdots, e_{n+1}] = e_{1}, \\
{[}e_{1}, e_{3}, \cdots, e_{n+1}] = e_{2}, \\
{[}e_{1}, \hat{e}_{2}, \hat{e}_{3}, \cdots, e_{n+2}] = b^{2}_{1, 3} e_{1}+b^{1}_{1, 3} e_{2}, \\
\cdots \cdots \cdots \cdots \cdots \cdots \cdots\\
{[}e_{1}, \hat{e}_{2}, \cdots, \hat{e}_{i}, \cdots, e_{n+2}] =
b^{2}_{1, i} e_{1}+b^{1}_{1, i} e_{2},
~ 3\leq i\leq n+1,\\
\cdots \cdots \cdots \cdots \cdots \cdots \cdots\\
{[}e_{1}, \hat{e}_{2}, \cdots, \hat{e}_{n+1}, \cdots, e_{n+2}] = b^{2}_{1, n+1} e_{1}+b^{1}_{1, n+1} e_{2}, \\
{[}e_{1}, e_{2}, \hat{e}_{3}, \hat{e}_{4}, \cdots, e_{n+2}] = b^{3}_{1, 4} e_{1}, \\
\cdots \cdots \cdots \cdots \cdots \cdots \cdots\\
{[}e_{1}, e_{2}, \hat{e}_{3}, \cdots, \hat{e}_{k}, \cdots, e_{n+2}]
= b^{3}_{1, k} e_{1},
~ 4\leq k\leq n+1,\\
\cdots \cdots \cdots \cdots \cdots \cdots \cdots\\
{[}e_{1}, e_{2}, \hat{e}_{3}, \cdots, \hat{e}_{n+1}, e_{n+2}] = b^{3}_{1, n+1} e_{1}, \\
{[}\hat{e}_{2}, e_3, \cdots, e_{n+2}] = b^{1}_{1, 2} e_{1}+b^{2}_{1, 2} e_{2},\\
\cdots \cdots \cdots \cdots \cdots \cdots \cdots\\
{[}e_{2}, \cdots, \hat{e}_{j}, \cdots, e_{n+2}] = b^{1}_{1, j}
e_{1}+b^{2}_{1, j} e_{2},
~ 2\leq j\leq n+1,\\
\cdots \cdots \cdots \cdots \cdots \cdots \cdots\\
{[}e_{2}, \cdots, \hat{e}_{n+1}, e_{n+2}] = b^{1}_{1, n+1} e_{1}+ b^{2}_{1, n+1} e_{2}.\\
\end{array}\right.
\end{array}$

\vspace{2mm}This implies $\dim A^1=2$. This is a contradiction.
Therefore, the case (3) is not realized. By  discussions similar to
the case (3), the cases $(4)$ and $(6)$ are not realized.

Now  substituting $e_1=[e_2, e_3, \cdots e_{n+1}]$ into the other
equations of (5) and imposing the Jacobi identities,  we get the
reduced form of (5) as follows

\vspace{2mm}\noindent$\begin{array}{l} (5)' ~
\left\{\begin{array}{l}
{[}e_{2}, \cdots, e_{n+1}] = e_{1}, \\
{[}e_{1}, \hat{e}_{2}, \hat{e}_{3}, \cdots, e_{n+2}] = (b^{2}_{1, 3} +b^{3}_{1, 2})e_{1}, \\
{[}e_{1}, \hat{e}_{2}, e_{3}, \hat{e}_{4}, \cdots, e_{n+2}] = b^{2}_{1, 4} e_{1}, \\
\cdots \cdots \cdots \cdots \cdots \cdots \cdots\\
{[}e_{1}, \hat{e}_{2}, e_3, \cdots, \hat{e}_{i}, \cdots, e_{n+2}] = b^{2}_{1, i} e_{1}, \\
\cdots \cdots \cdots \cdots \cdots \cdots \cdots\\
{[}e_{1}, \hat{e}_{2}, e_3, \cdots, \hat{e}_{n+1}, e_{n+2}] = b^{2}_{1, n+1} e_{1}, \\
{[}e_{1}, e_{2}, \hat{e}_{3}, \hat{e}_{4}, \cdots, e_{n+2}] = b^{3}_{1, 4} e_{1}, \\
\cdots \cdots \cdots \cdots \cdots \cdots \cdots\\
{[}e_{1}, e_{2}, \hat{e}_{3},  \cdots, \hat{e}_{j}, \cdots, e_{n+2}] = b^{3}_{1, j} e_{1}, \\
\cdots \cdots \cdots \cdots \cdots \cdots \cdots\\
{[}e_{1}, e_{2}, \hat{e}_{3}, \cdots, \hat{e}_{n+1}, e_{n+2}] = b^{3}_{1, n+1} e_{1}, \\
{[}\hat{e}_{2}, e_{3},  \cdots, e_{n+2}] = b^{2}_{1, 2} e_{2}+b^{3}_{1, 2} e_{3}, \\
\cdots \cdots \cdots \cdots \cdots \cdots \cdots\\
{[}e_{2}, \cdots, \hat{e}_{k}, \cdots, e_{n+2}] = b^{2}_{1, k} e_{2}+b^{3}_{1, k} e_{3},\\
\cdots \cdots \cdots \cdots \cdots \cdots \cdots\\
{[}e_{2}, \cdots, \hat{e}_{n+1}, e_{n+2}] = b^{2}_{1, n+1} e_{2}+b^{3}_{1, n+1} e_{3}. \\
\end{array}\right.
\end{array} 4\leq i, j\leq n+1, 2\leq k\leq n+1.$

\vspace{1mm}We claim  $b^2_{1, l}=0$ for $l>3$. If there exists $l
>3$ such that $b^{2}_{1, l}\neq0,$ replacing
$e_{l}+\sum\limits_{m=4}^{n+1}\frac{b^{2}_{1, m}}{b^{2}_{1, l}}
e_{m}$ for $e_{l}$ in $(5)'$  and using the Jacobi identities for $
\{\frac{1}{b^{2}_{1, l}}([e_{2}, \cdots,$ $ \hat{e}_{l}, \cdots,$
$e_{n+2}]+b^{3}_{1, l} e_{3}),$ $ e_{3}, \cdots, \hat{e}_{t},
\cdots, e_{n+2}\},$  ~$4\leq t\leq n+1,$ ~$t\neq l,$
 we get

\vspace{2mm}\noindent \noindent $\begin{array}{l}
\left\{\begin{array}{l}
{[}e_{2}, \cdots, e_{n+1}] = e_{1}, \\
{[}e_{1}, \hat{e}_{2}, \hat{e}_{3},  \cdots, e_{n+2}] = (b^{2}_{1, 3} +b^{3}_{1, 2})e_{1}, \\
{[}e_{1}, \hat{e}_{2}, e_{3}, \hat{e}_{4}, \cdots, e_{n+2}] = b^{2}_{1, 4} e_{1}, \\
\cdots \cdots \cdots \cdots \cdots \cdots \cdots\\
{[}e_{1}, \hat{e}_{2}, e_3, \cdots, \hat{e}_{i}, \cdots, e_{n+2}] = b^{2}_{1, i} e_{1}, \\
\cdots \cdots \cdots \cdots \cdots \cdots \cdots\\
{[}e_{1}, \hat{e}_{2}, e_3,  \cdots, \hat{e}_{n+1}, e_{n+2}] = b^{2}_{1, n+1} e_{1}, \\
{[}e_{1}, e_{2}, \hat{e}_{3}, \hat{e}_{4}, \cdots, e_{n+2}] = b^{3}_{1, 4} e_{1}, \\
\cdots \cdots \cdots \cdots \cdots \cdots \cdots\\
{[}e_{1}, e_{2}, \hat{e}_{3}, \cdots, \hat{e}_{j}, \cdots, e_{n+2}] = b^{3}_{1, j} e_{1}, \\
\cdots \cdots \cdots \cdots \cdots \cdots \cdots\\
{[}e_{1}, e_{2}, \hat{e}_{3}, \cdots, \hat{e}_{n+1}, e_{n+2}] = b^{3}_{1, n+1} e_{1}, \\
{[}\hat{e}_{2}, e_3, \cdots, e_{n+2}] = b^{2}_{1, 2} (e_{2}+
\frac{b^{3}_{1, l}}{b^{2}_{1, l}} e_{3}),\\
\cdots \cdots \cdots \cdots \cdots \cdots \cdots\\
{[}e_{2}, \cdots, \hat{e}_{k}, \cdots, e_{n+2}] = b^{2}_{1, k}(e_{2}+
\frac{b^{3}_{1, l}}{b^{2}_{1, l}} e_{3}), \\
\cdots \cdots \cdots \cdots \cdots \cdots \cdots\\
{[}e_{2}, \cdots, \hat{e}_{n+1}, e_{n+2}] = b^{2}_{1, n+1} (e_{2}+
\frac{b^{3}_{1, l}}{b^{2}_{1, l}} e_{3});
\end{array}\right.
\end{array} 4\leq i, j \leq n+1, 2\leq k\leq n+1.$

\noindent This implies  $\dim A^1=2$. This is a contradiction.
Therefore $b^{2}_{1, l}=0$ for $l > 3$. Similarly,
 we have $b^{3}_{1, l}=0$ for $l > 3$. Therefore, $(5)'$ is of
the form

\vspace{2mm}\noindent $\begin{array}{l} (5)'' ~
\left\{\begin{array}{l}
{[}e_{2}, \cdots, e_{n+1}] = e_{1}, \\
{[}e_{1}, e_{4}, \cdots, e_{n+2}] = (b^{2}_{1, 3} +b^{3}_{1, 2})e_{1}, \\
{[}e_{2}, e_{4}, \cdots, e_{n+2}] = b^{2}_{1, 3} e_{2}+b^{3}_{1, 3} e_{3}, \\
{[}e_{3}, \cdots, e_{n+2}] = b^{2}_{1, 2} e_{2}+b^{3}_{1, 2} e_{3}.
\end{array}\right.
\end{array}$

\vspace{1mm}\noindent If $b^{2}_{1, 2}\neq 0$, replacing $e_2$ by
$e_{2}+\frac{b^{2}_{1, 3}}{b^{2}_{1, 2}}e_3$ in $(5)''$, (5) is
isomorphic to

\vspace{2mm}\noindent $\begin{array}{l} (5)'''
~\left\{\begin{array}{l}
{[}e_{2}, \cdots, e_{n+1}] = e_{1}, \\
{[}e_{1}, e_{4}, \cdots, e_{n+2}] = (b^{2}_{1, 3} +b^{3}_{1, 2})e_{1}, \\
{[}e_{2}, e_{4}, \cdots, e_{n+2}] = b^{2}_{1, 3} e_{3}, \\
{[}e_{3}, \cdots, e_{n+2}] = b^{2}_{1, 2} e_{2}+(b^{3}_{1,
2}+b^{2}_{1, 3}) e_{3}.
\end{array}\right.
\end{array}$

\vspace{1mm}\noindent When $b^{3}_{1, 2}=b^{2}_{1, 3}$, since $\dim
A^1=3$, $b^{2}_{1, 3}\neq 0.$  Taking a linear transformation of the
basis $e_1, $ $\cdots, $ $e_{n+2}$ by replacing $e_1,$ $e_3$ and
$e_{n+2}$ by $\sqrt{\frac{b_{1, 3}^2}{b_{1, 2}^2}}e_1$,
$\sqrt{\frac{b_{1, 3}^2}{b_{1, 2}^2}}e_3$ and $\sqrt{\frac{1}{b_{1,
2}^2b_{1, 3}^2}}e_{n+2}$ respectively, we get that
  (5) is isomorphic to
$\begin{array}{l} (d^{4}) ~ \left\{\begin{array}{l}
{[}e_{2}, \cdots, e_{n+1}] = e_{1}, \\
{[}e_{2}, e_{4}, \cdots, e_{n+2}] = e_{3}, \\
{[}e_{3}, \cdots, e_{n+2}] = e_{2}.
\end{array}\right.
\end{array} $

\vspace{1mm} When $b^{3}_{1, 2}\neq b^{2}_{1, 3}$, we also have
$b^{2}_{1, 3}\neq 0$. Substituting $\frac{1}{b^{2}_{1, 3}}e_1$ for
$e_{1}$, $\frac{1}{b^{2}_{1, 3}}e_2$ for $e_{2}$ and
$\frac{1}{b^{2}_{1, 3} +b^{3}_{1, 2}}e_{n+2}$ for $e_{n+2}$ in
$(5)'''$, we have
 $\begin{array}{l} (d^{6})
~\left\{\begin{array}{l}
{[}e_{2}, \cdots, e_{n+1}] = e_{1}, \\
{[}e_{1}, e_{4}, \cdots, e_{n+2}] = e_{1}, \\
{[}e_{2}, e_{4}, \cdots, e_{n+2}] = e_{3}, \\
{[}e_{3}, \cdots, e_{n+2}] = e_{2}+\gamma e_{3},\end{array}\right.
\end{array}~ \gamma\in F, \gamma\neq 0.$

\vspace{1mm} If $b^{2}_{1, 2}=0$, then $b^{2}_{1, 3}\neq0$. Taking a
suitable linear transformation of the basis $e_1, \cdots,$ $
e_{n+2},$ we have $(5)''$ is isomorphic to $\begin{array}{l}
(d^{5})~ \left\{\begin{array}{l}
{[}e_{2}, \cdots, e_{n+1}] = e_{1}, \\
{[}e_{2}, e_{4}, \cdots, e_{n+2}] = e_{3}, \\
{[}e_{3}, \cdots, e_{n+2}] = e_{2}.
\end{array}\right.
\end{array}$

 \vspace{1mm} By Lemma 2.1 an $n$-Lie algebra of the case $(d^6)$
has a unique nonabelian ideal $I=Fe_1+\cdots+Fe_{n+1}$ (up to an
isomorphism) of  codimension  $1$ with $I^1=Z(I)$. Suppose $A_1$ and
$A_2$ are isomorphic $n$-Lie algebras of the case $(d^6)$ with
nonzero coefficients $\gamma$ and $\gamma'$, and let $I_1$ and $I_2$
denote the nonabelian ideals of codimension  $1$ described above.
Let $\sigma: A_1\rightarrow A_2$ be an $n$-Lie isomorphism from
$A_1$ to $A_2$, then $\sigma(I_1)=\sigma(I_2)$. And the transition
matrix of $\sigma$ from $A_1$ to $A_2$ is of the form
$$
T=\left(
      \begin{array}{ccccccc}
\lambda_{1, 1}&\lambda_{1, 2}&\lambda_{1, 3}&0&\cdots&0&\lambda_{1, n+2}\\
0&\lambda_{2, 2}&\lambda_{2, 3}&0&\cdots&0&\lambda_{2, n+2}\\
0&\lambda_{3, 2}&\lambda_{3, 3}&0&\cdots&0&\lambda_{3, n+2}\\
0&0&0&1&\cdots&0&\lambda_{4, n+2}\\
\vdots&\vdots &\vdots &\vdots&\cdots&\vdots&\vdots\\
0&0&0&0&\cdots&1&\lambda_{n+1, n+2}\\
0&0&0&0&\cdots&0&\lambda_{n+2, n+2}
\end{array}
    \right),~ \mbox{det}~T \neq 0,
    $$
    that is $\sigma(e_1, \cdots, e_{n+2})=(e^1_1, \cdots,  e^1_{n+2})T,$
where $e_1, \cdots, e_{n+2}$ and $e^1_1, \cdots, e^1_{n+2}$ are
basis of $ A_1$ and $A_2$ respectively, and $A_1$ and $A_2$ has the
multiplication table $(d^6)$ in the basis $e_1, \cdots, e_{n+2}$ and
$e^1_1, \cdots, e^1_{n+2}$. By direct computation of
$$\sigma([e_1, \cdots, \hat{e}_i, \cdots,
\hat{e}_j, \cdots, e_{n+2}])=[\sigma(e_1), \cdots,
\sigma(\hat{e}_i), \cdots, \sigma(\hat{e}_j), \cdots,
\sigma(e_{n+2)})],$$ where $1\leq i, j\leq n+2, i\neq j$, we get
$\gamma=\gamma'.$ Therefore  $n$-Lie algebras of the case $(d^6)$
with coefficients $\gamma$ and $\gamma'$ are isomorphic if and only
if $\gamma=\gamma'$.

Lastly we discuss the case (7). It follows by a simple computation
that there does not exist any nonabelian proper subalgebra of $A$
containing $A^1.$ Then the
 the multiplication of $A$ is completely determined
by the left multiplication ad$(e_{4},\cdots, e_{n+2})$. And
ad$(e_{4},\cdots, e_{n+2})|_{A^1}$ is  nonsingular since $\dim
A^1=3$. So we can choose a  basis $e_1, e_2, e_3$ of $A^1$ such that
the multiplication of $A$ in the basis $e_1, \cdots, e_{n+2}$ has
the following  possibilities

\vspace{2mm}\noindent$\begin{array}{l} (d^{7})' ~
\left\{\begin{array}{l}
{[}e_{1}, e_{4}, \cdots, e_{n+2}]=\beta_{1}e_{1}, \\
{[}e_{2}, e_{4}, \cdots, e_{n+2}]=\beta_{2}e_{2}, \\
{[}e_{3}, e_4, \cdots, e_{n+2}]=\beta_{3}e_{3},
\end{array}\right.
\end{array} ~\beta_{i}\in F, \beta_i\neq0, ~i=1, 2, 3;
$

\vspace{2mm}\noindent$
\begin{array}{l}
(d^{8})' ~ \left\{\begin{array}{l}
{[}e_{1}, e_{4}, \cdots, e_{n+2}]=\alpha e_{1}+e_{2}, \\
{[}e_{2}, e_{4}, \cdots, e_{n+2}]=\alpha e_{2}+e_{3}, \\
{[}e_{3}, e_4, \cdots, e_{n+2}]=\alpha e_{3}, \\
\end{array}\right. \alpha\in F, \alpha\neq 0;
\end{array}
$

\vspace{2mm}\noindent$\begin{array}{l} (d^{9})'  ~
\left\{\begin{array}{l}
{[}e_{1}, e_{4}, \cdots, e_{n+2}]=\gamma _{1} e_{1}+e_{2}, \\
{[}e_{2}, e_{4}, \cdots, e_{n+2}]=\gamma_{1} e_{2}, \\
{[}e_{3}, e_4, \cdots, e_{n+2}]=\gamma_{2} e_{3};
\end{array}\right.
\end{array} ~\gamma_j\in F, \gamma_j\neq 0, ~j=1, 2.$

\vspace{1mm}If we fix $e_5, \cdots, e_{n+2}$ in the $n$-ary
multiplication of $A$, we get a solvable Lie algebra $A_1=A$ ( as
vector spaces) with the Lie product  $[ , ]_1$
 as follows
 $$
 [x, y]_1=[x, y, e_5, \cdots, e_{n+2}], ~ x, y\in A_1.
 $$
Then the multiplication tables of $A_1$  with respect to
$(d^7)', (d^8),' (d^{9})'$  are

\vspace{2mm}\noindent$\begin{array}{l} (d^{7})'' ~
\left\{\begin{array}{l}
{[}e_{1}, e_{4}]_1=\beta_{1}e_{1}, \\
{[}e_{2}, e_{4}]_1=\beta_{2}e_{2}, \\
{[}e_{3}, e_{4}]_1=\beta_{3}e_{3},
\end{array}\right.
\end{array} ~\beta_{i}\in F, \beta_i\neq0, 1=1, 2, 3;
$

\vspace{2mm}\noindent$
\begin{array}{l}(d^{8})'' ~ \left\{\begin{array}{l}
{[}e_{1}, e_{4}]_1=\alpha e_{1}+e_{2}, \\
{[}e_{2}, e_{4}]_1=\alpha e_{2}+e_{3}, \\
{[}e_{3}, e_{4}]_1=\alpha  e_{3}, \\
\end{array}\right.
\end{array} \alpha\in F, \alpha\neq 0;
$

\vspace{2mm}\noindent$
\begin{array}{l}
(d^{9})''  ~ \left\{\begin{array}{l}
{[}e_{1}, e_{4}]_1=\gamma _{1}e_{1}+e_{2}, \\
{[}e_{2}, e_{4}]_1=\gamma_{1}e_{2}, \\
{[}e_{3}, e_{4}]=\gamma_{2}e_{3};
\end{array}\right.
\end{array} ~\gamma_j\in F, \gamma_j\neq 0, j=1, 2.
$

\vspace{1mm}This implies that $A_1$ can be decomposed into the
direct sum of ideals $Z(A_1)$ and $B$, where the center
$Z(A_1)=Fe_5+\cdots+Fe_{n+2}$ and the ideal $B=Fe_1+Fe_2+Fe_3+Fe_4$.
By the classification of $4$-dimensional solvable Lie algebras [8],
we get that one and only one of following possibilities holds up to
isomorphism

\vspace{2mm}\noindent $\begin{array}{l} (d^{7})~
\left\{\begin{array}{l}
{[}e_{1}, e_{4}, \cdots, e_{n+2}]=e_{1},\\
{[}e_{2}, e_{4}, \cdots, e_{n+2}]= e_{3},\\
{[}e_{3}, e_4, \cdots, e_{n+2}]= \beta e_2+(1+\beta)e_{3}, ~
\beta\in F, \beta\neq 0, 1;
\end{array}\right.
\end{array}$

\vspace{2mm}\noindent$ \begin{array}{l} (d^{8})~
\left\{\begin{array}{l}
{[}e_{1}, e_{4}, \cdots, e_{n+2}]=e_{1},\\
{[}e_{2}, e_{4}, \cdots, e_{n+2}]=e_{2},\\
{[}e_{3}, e_{4}, \cdots, e_{n+2}]=e_{3};
\end{array}\right.
\end{array}
$

\vspace{2mm}\noindent $\begin{array}{l} (d^{9})~
\left\{\begin{array}{l}
{[}e_{1}, e_{4}, \cdots, e_{n+2}]=e_{2},\\
{[}e_{2}, e_{4}, \cdots, e_{n+2}]= e_{3},\\
{[}e_{3}, e_{4}, \cdots, e_{n+2}]= se_1+te_2+ue_{3}, ~s, t, u\in
F,~s\neq 0.
\end{array}\right.
\end{array}
$

\vspace{1mm} \noindent And $(d^i)$ is not isomorphic to $(d^j)$ when
$i\neq j$ for $ 7\leq i, ~j\leq 9.$ And the $n$-Lie algebras
corresponding to the case $(d^7)$ with coefficients $\beta$ and
$\beta'$ are isomorphic if and only if $\beta=\beta'.$  We also have
that the $n$-Lie algebras corresponding to the case $(d^{9})$ with
coefficients  $s, t, u$ and $s', t', u'$ are isomorphic  if and only
if there exists a nonzero element $r\in F$ such that
$$
s=r^3 s', ~t=r^2 t', ~u=ru', ~s, s', t, t', u, u' \in F.
$$

It is evident that $(d^{7}), (d^{8}), (d^{9})$ are not isomorphic
 to other cases since
$(d^{7}),$ $ (d^{8}), $ $(d^{9})$ have  no nonabelian proper
subalgebras containing $A^1$. $(d^{1})$ and $(d^{3})$ are not
isomorphic to any cases of $(d^{2}), (d^{4}), (d^{5})$ and $(d^{6})$
since $(d^{1})$ and $(d^{3})$ are decomposable. And thanks to Lemma
2.1 $(d^{1})$ is not isomorphic to $(d^{3})$. Since $(d^{4})$ and
$(d^{5})$ have a non-trivial center, $(d^{4})$ and $(d^{5})$ are not
isomorphic to $(d^{2})$ and $(d^{6})$.

That $(d^4)$ is not isomorphic to $(d^5)$ follows at once from the
fact that the maximum of dimensions of  toral subalgebras of $(d^4)$
is $n-1$ but the maximum of dimensions of  toral subalgebras of
$(d^5)$ is $n$.

Now we study the cases $(d^2)$ and $(d^6)$. It is not difficult to
see that two cases have  $(n-1)$-dimensional Toral subalgebras. In
the case of $(d^6)$, there exists an $(n-1)$-dimensional Toral
subalgebra $H=Fe_4+\cdots+Fe_{n+2}$ such that the derived algebra
$A^1$ is a completely reducible $H$-module with nontrivial action.
But there does not exist such $(n-1)$-dimensional Toral subalgebra
in the case of $(d^2)$. Therefore, $(d^2)$ is not isomorphic to
$(d^6).$

5.\quad  If $\dim A^{1}=r$, $r$ is even and $r\geq 4$. Suppose
$A^1=F e_1+F e_2+\cdots+F e_r$. Then the  multiplication falls into
one of the following cases

\vspace{2mm}\noindent$\begin{array}{l} (1)~ \left\{\begin{array}{l}
{[}\hat{e}_{1}, e_{2}, \cdots, e_{n+1}] = e_{1}, \\
{[} e_{1}, \hat{e}_2, \cdots, e_{n+1}] = e_{2}, \\
\cdots \cdots \cdots \cdots \cdots \cdots \cdots\\
{[}e_{1}, \cdots, \hat{e}_{p}, \cdots, e_{n+1}] = e_{p}, \\
{[}e_{1}, \cdots, \hat{e}_{p+1}, \cdots, e_{n+1}] = e_{r}, \\
{[}e_{1}, \cdots, \hat{e}_{p+2}, \cdots, e_{n+1}] = e_{r-1}, \\
\cdots \cdots \cdots \cdots \cdots \cdots \cdots\\
{[}e_{1}, \cdots, \hat{e}_{p+q}, \cdots, e_{n+1}] = e_{p+1}, ~~2\leq q \leq r, ~p+q=r, ~q~\mbox{is even},\\
{[}e_{1}, \cdots, \hat{e}_{i}, \cdots, \hat{e}_{j}, \cdots, e_{n+2}]
=\sum\limits_{k=1}^{r}b^{k}_{i, j} e_{k};
\end{array}\right.
\end{array}$

\vspace{2mm}\noindent$\begin{array}{l} (2)~ \left\{\begin{array}{l}
{[} \hat{e}_1, e_{2}, \cdots, e_{n+1}] = e_{1}, \\
\cdots \cdots \cdots \cdots \cdots \cdots \cdots\\
{[}e_{1}, \cdots, \hat{e}_{k}, \cdots, e_{n+1}] = e_{k}, ~1\leq k\leq r,\\
\cdots \cdots \cdots \cdots \cdots \cdots \cdots\\
{[}e_{1}, \cdots, \hat{e}_{r}, \cdots, e_{n+1}] = e_{r},\\
{[}e_{1},  \cdots, \hat{e}_{i}, \cdots, \hat{e}_{j}, \cdots,
e_{n+2}] =\sum\limits_{k=1}^{r}b^{k}_{i, j} e_{k};
\end{array}\right.
\end{array}$

\vspace{2mm}\noindent$\begin{array}{l} (3)~ \left\{\begin{array}{l}
{[} \hat{e}_1, e_{2},  \cdots, e_{n+1}] = e_{1}, \\
\cdots \cdots \cdots \cdots \cdots \cdots \cdots\\
{[} e_{1}, \cdots, \hat{e}_k, \cdots, e_{n+1}] = e_{k}, ~ 1\leq k \leq p, \\
\cdots \cdots \cdots \cdots \cdots \cdots \cdots\\
{[}e_{1}, \cdots, \hat{e}_{p}, \cdots, e_{n+1}] = e_{p}, \\
{[}e_{1}, \cdots, \hat{e}_{p+1}, \cdots, e_{n+1}] = e_{r-1}, \\
\cdots \cdots \cdots \cdots \cdots \cdots \cdots\\
{[}e_{1}, \cdots, \hat{e}_{p+l}, \cdots, e_{n+1}] = e_{r-l}, ~1\leq l\leq q,\\
\cdots \cdots \cdots \cdots \cdots \cdots \cdots\\
{[}e_{1}, \cdots, \hat{e}_{p+q}, \cdots, e_{n+1}] = e_{p+1},~ 2\leq q < r,~p+q=r-1,~q~\mbox{is even},\\
{[}e_{1}, \cdots, \hat{e}_{i}, \cdots, \hat{e}_{j}, \cdots, e_{n+2}]
=\sum\limits_{k=1}^{r}b^{k}_{i, j} e_{k};
\end{array}\right.
\end{array}$

\vspace{2mm}\noindent$\begin{array}{l} (4)~  \left\{\begin{array}{l}
{[}\hat{e}_1, e_{2}, \cdots, e_{n+1}] = e_{1}, \\
\cdots \cdots \cdots \cdots \cdots \cdots \cdots\\
{[} e_{1},  \cdots, \hat{e}_k, \cdots, e_{n+1}] = e_{k}, ~1\leq k\leq r-1, \\
\cdots \cdots \cdots \cdots \cdots \cdots \cdots\\
{[}e_{1}, \cdots, \hat{e}_{r-1}, \cdots, e_{n+1}] = e_{r-1},\\
{[}e_{1}, \cdots, \hat{e}_{i}, \cdots, \hat{e}_{j}, \cdots, e_{n+2}]
=\sum\limits_{k=1}^{r}b^{k}_{i, j} e_{k};
\end{array}\right.
\end{array}$

\vspace{2mm}\noindent$~~\cdots \cdots \cdots \cdots \cdots \cdots
\cdots\cdots \cdots \cdots \cdots \cdots \cdots \cdots$

\vspace{2mm}\noindent$\begin{array}{l} (2r-3)~
\left\{\begin{array}{l}
{[}e_{2}, \cdots, e_{n+1}] = e_{1}, \\
{[}e_{1}, e_{3}, \cdots, e_{n+1}] = e_{2}, \\
{[}e_{1}, \cdots, \hat{e}_{i}, \cdots, \hat{e}_{j}, \cdots, e_{n+2}]
=\sum\limits_{k=1}^{r}b^{k}_{i, j} e_{k};
\end{array}\right.
\end{array} $

\vspace{2mm}\noindent$\begin{array}{l} (2r-2)~
\left\{\begin{array}{l}
{[}e_{2}, \cdots, e_{n+1}] = e_{1}+\alpha e_{2},\\
{[}e_{1}, e_{3}, \cdots, e_{n+1}] = e_{2}, \\
{[}e_{1}, \cdots, \hat{e}_{i}, \cdots, \hat{e}_{j}, \cdots, e_{n+2}]
=\sum\limits_{k=1}^{r}b^{k}_{i, j} e_{k};
\end{array}\right.
\end{array} $

\vspace{2mm}\noindent$\begin{array}{l} (2r-1)~
\left\{\begin{array}{l}
{[}e_{1}, \cdots, e_{n}] = e_{1},\\
{[}e_{1}, \cdots, \hat{e}_{i}, \cdots, \hat{e}_{j}, \cdots, e_{n+2}]
=\sum\limits_{k=1}^{r}b^{k}_{i, j} e_{k};
\end{array}\right.
\end{array} $

\vspace{2mm}\noindent$\begin{array}{l} (2r)~ \left\{\begin{array}{l}
{[}e_{2}, \cdots, e_{n+1}] = e_{1},\\
{[}e_{1}, \cdots, \hat{e}_{i}, \cdots, \hat{e}_{j}, \cdots, e_{n+2}]
=\sum\limits_{k=1}^{r}b^{k}_{i, j} e_{k};
\end{array}\right.
\end{array} $

\vspace{2mm}\noindent$\begin{array}{l} (2r+1)~
\left\{\begin{array}{l} {[}\hat{e}_{1}, \hat{e}_{2}, \cdots,
e_{r+1}, \cdots, e_{n+2}]
=b^{1}_{1, 2} e_{1}+b^{2}_{1, 2} e_{2}+\cdots+b^{r}_{1, 2} e_{r},\\
\cdots \cdots \cdots \cdots \cdots \cdots \cdots\\
{[}e_{1}, \cdots, \hat{e}_{k}, \cdots, \hat{e}_{l}, \cdots, e_{r+1},
\cdots, e_{n+2}]
=b^{1}_{k, l} e_{1}+b^{2}_{k, l} e_{2}+\cdots+b^{r}_{k, l} e_{r},\\
\cdots \cdots \cdots \cdots \cdots \cdots \cdots\\
{[}e_{1}, \cdots, \hat{e}_{r-1}, \hat{e}_{r}, e_{r+1}, \cdots,
e_{n+2}] =b^{1}_{r-1, r} e_{1}+b^{2}_{r-1, r}
e_{2}+\cdots+b^{r}_{r-1, r} e_{r};
\end{array}\right.
\end{array}$

\vspace{2mm}\noindent where $1\leq i\neq j \leq n+1$, $1\leq k\neq l
\leq r$.

Firstly, we study the table (1) in the case of $p=0$. Then (1) is of
the form

\vspace{2mm}\noindent $\begin{array}{l}  \left\{\begin{array}{l}
{[} \hat{e}_{1}, e_{2}, \cdots, e_{n+1}] = e_{r},\\
\cdots \cdots \cdots \cdots \cdots \cdots \cdots\\
{[}e_{1}, \cdots, \hat{e}_{m},\cdots, e_{n+1}] = e_{r-m+1}, ~ 1\leq m\leq r,\\
\cdots \cdots \cdots \cdots \cdots \cdots \cdots\\
{[}e_{1}, \cdots, \hat{e}_{r}, \cdots, e_{n+1}] = e_{1}, \\
{[}e_{1}, \cdots, \hat{e}_{i}, \cdots, \hat{e}_{j},\cdots, e_{n+2}]
=\sum\limits_{k=1}^{r}b^{k}_{i, j} e_{k}, ~1\leq i\neq j \leq n+1.
\end{array}\right.
\end{array}
$

\vspace{2mm}\noindent It is similar to the cases $r\leq 3$. By
imposing the Jacobi identities on the above table for

\vspace{2mm}$\{e_{1},$  $e_2$, $e_3,$ $\cdots,$ $ \hat{e}_{i},$ $
\cdots, $ $\hat{e}_{j},$ $ \cdots, $ $e_{n+2}\}$, $3 < i \neq j \leq
r$,

\vspace{2mm}$\{e_{1},$ $e_{3},$ $\cdots,$ $\hat{e}_{j},$ $\cdots,$ $
e_{n+2}\}$, ~$3< j \leq r,$

\vspace{2mm}$\{e_{1},$ $e_{2},$ $\hat{e}_{3},$ $\cdots,$
$\hat{e}_{j},$ $\cdots, $ $e_{n+2}\}$, ~$3< j\leq r,$

\vspace{2mm}$\{e_{1},$ $e_{4},$ $\cdots,$ $e_{n+2}\}$, ~$\{e_{3},$
$\cdots,$ $e_{n+2}\}$, ~$\{e_{2},$ $e_{4},$ $\cdots,$ $e_{n+2}\}$,

\vspace{2mm}$\{e_{2},$ $e_3,$ $\cdots,$ $\hat{e}_{j},$ $\cdots,$ $
e_{n+2}\},$ ~$ 3< j\leq r$,

\vspace{2mm}$\{e_{1},$ $ e_2$, $\cdots,$ $\hat{e}_{i},$ $\cdots,$
$\hat{e}_{j},$ $\cdots,$ $ e_{n+2}\}$, $3\leq i\leq r< j\leq n+1$,

\vspace{2mm}$\{e_{1},$ $e_{3},$ $\cdots,$ $\hat{e}_{j},$ $\cdots, $
$e_{n+2}\}$, ~$r< j\leq n+1,$ and

\vspace{2mm} $\{e_{2}, $ $\cdots,$ $\hat{e}_{j},$ $\cdots,$ $
e_{n+2}\}$, ~$r< j\leq n+1$, we get

\vspace{2mm}\noindent $\begin{array}{l} (1)'
~\left\{\begin{array}{l}
{[}\hat{e}_1, e_{2}, \cdots, e_{n+1}] = e_{r}, \\
\cdots \cdots \cdots \cdots \cdots \cdots \cdots\\
{[}e_{1}, \cdots, \hat{e}_{m},\cdots, e_{n+1}] = e_{r-m+1}, ~ 1\leq m\leq r,\\
\cdots \cdots \cdots \cdots \cdots \cdots \cdots\\
{[}e_{1}, \cdots, \hat{e}_{r}, \cdots, e_{n+1}] = e_{1}, \\
{[}e_{3}, \cdots, e_{n+2}] = b^{r-3}_{2, 4}e_{r}+b^{r-2}_{1, 3}e_{r-1}, \\
{[}e_{2}, \hat{e}_3, \cdots, e_{n+2}] = b^{r-3}_{3, 4}e_{r}+b^{r-2}_{1, 3}e_{r-2}, \\
\cdots \cdots \cdots \cdots \cdots \cdots \cdots\\
{[}e_{2}, \cdots, \hat{e}_{j},\cdots, e_{n+2}] = b^{r-2}_{3, j}e_{r}+b^{r-2}_{1, 3}e_{r-j+1}, ~3\leq j\leq r,\\
\cdots \cdots \cdots \cdots \cdots \cdots \cdots\\
{[}e_{2}, \cdots, \hat{e}_{r}, \cdots, e_{n+2}] = b^{r-2}_{3, r}e_{r}+b^{r-2}_{1, 3}e_{1}, \\
{[}e_{1}, e_4, \cdots, e_{n+2}] = b^{r-3}_{3, 4}e_{r-1}+b^{r-3}_{2, 4}e_{r-2}, \\
{[}e_{1}, \hat{e}_2, e_3, \hat{e}_{4}, \cdots, e_{n+2}] = b^{r-2}_{3, 4}e_{r-1}+b^{r-3}_{2, 4}e_{r-3}, \\
\cdots \cdots \cdots \cdots \cdots \cdots \cdots\\
{[}e_{1}, \hat{e}_2, e_3, \cdots, \hat{e}_{i}, \cdots, e_{n+2}] =
b^{r-2}_{3, i}e_{r-1}+b^{r-3}_{2, 4}e_{r-i+1}, ~4\leq i\leq r, \\
\cdots \cdots \cdots \cdots \cdots \cdots \cdots\\
{[}e_{1}, \hat{e}_2, e_3, \cdots, \hat{e}_{r}, \cdots, e_{n+2}] =
b^{r-2}_{3, r}e_{r-1}+b^{r-3}_{2, 4}e_{1}, \\
{[}e_{1}, e_{2}, \hat{e}_3, \hat{e}_4,  \cdots, e_{n+2}] = b^{r-2}_{3, 4}e_{r-2}+b^{r-1}_{2, 3}e_{r-3}, \\
\cdots \cdots \cdots \cdots \cdots \cdots \cdots\\
{[}e_{1}, e_2, \cdots, \hat{e}_{l}, \cdots, \hat{e}_{l'}, \cdots,
e_{n+2}] =b^{r-2}_{3, l'}e_{r-l+1}+b^{r-1}_{2, l}e_{r-l'+1}, ~ 3\leq l< l'\leq r,\\
\cdots \cdots \cdots \cdots \cdots \cdots \cdots\\
{[}e_{1}, e_2, \cdots, \hat{e}_{r-1}, \hat{e}_{r}, \cdots,
e_{n+2}] =b^{r-2}_{3, r}e_{2}+b^{r-1}_{2, r-1}e_{1} \\
{[}\hat{e}_1, e_2, \cdots, e_{r}, \hat{e}_{r+1}, \cdots, e_{n+2}] = b^{r-1}_{2, r+1}e_{r}, \\
\cdots \cdots \cdots \cdots \cdots \cdots \cdots\\
{[}e_{1}, \cdots, \hat{e}_{k}, \cdots, \hat{e}_{k'}, \cdots,
e_{n+2}] =b^{r-1}_{2, k'}e_{r-k+1}, ~ 1\leq k\leq r, ~ r< k'\leq n+1,\\
\cdots \cdots \cdots \cdots \cdots \cdots \cdots\\
{[}e_{1}, \cdots, \hat{e}_{r}, \cdots, \hat{e}_{n+1}, e_{n+2}]
=b^{r-1}_{2, n+1}e_{1}.
\end{array}\right.
\end{array} $

 \vspace{1mm}\noindent Substituting $e_{n+2}+b^{r-2}_{1, 3} e_{1}+b^{r-3}_{2, 4}
e_{2}+b^{r-3}_{3, 4}e_{3}+\sum\limits_{s=4}^{r}b^{r-2}_{3, s}
e_{s}+\sum\limits_{t=r+1}^{n+1}b^{r-1}_{2, t} e_{t} $ for $e_{n+2}$
in $(1)'$, we obtain  $(e^1)$ in the case of $q=r$ $ (p=0)$:

\vspace{2mm}\noindent $\begin{array}{l} \left\{\begin{array}{l}
{[}\hat{e}_1, e_{2}, \cdots, e_{n+1}] = e_{r}, \\
\cdots \cdots \cdots \cdots \cdots \cdots \cdots\\
{[}e_{1}, \cdots, \hat{e}_{i}, \cdots, e_{n+1}] = e_{r-i+1},  \\
\cdots \cdots \cdots \cdots \cdots \cdots \cdots\\
{[}e_{1}, \cdots, \hat{e}_{r}, \cdots, e_{n+1}] = e_{1};
\end{array}\right.
\end{array} 1\leq i\leq r.$

\vspace{1mm} Similarly, when the even number $q$ satisfying $q <r $
or $p > 0$, the cases (1) and (2) are isomorphic to $(e^1)$ and
$(e^2)$ respectively.

Secondly, by the similar discussion to the case (1),  we obtain that
$\dim A^1 \leq r-1$ for the cases $ (3),$ $ (4),$ $ \cdots,$ $
(2r-1)$ and $(2r+1)$. These are contradictions. Therefore the cases
$ (3),$ $ (4),$ $ \cdots,$ $ (2r-1)$ and $(2r+1)$ are not realized.

Now we study $(2r)$. By substituting
$e_{n+2}+\sum\limits_{j=2}^{n+1}b^{1}_{1, j} e_{j}$ for $e_{n+2}$ in
$(2r)$, and by formula

${[}e_{1}, \cdots, \hat{e}_{i}, \cdots, \hat{e}_{j}, \cdots,
e_{n+2}] =[{[}e_{2}, \cdots, e_{n+1}], e_{2}, \cdots, \hat{e}_{i},
\cdots, \hat{e}_{j}, \cdots, e_{n+2}]$

$=\sum\limits_{k=1}^{r}b^{k}_{i, j} e_{k}=(b_{1, j}^{i}+b_{1,
i}^{j})e_{1}, ~2\leq i\neq j\leq n+1,$
\\we have that $(2r)$ is
isomorphic to

\vspace{2mm}\noindent $\begin{array}{l} (2r)' ~
\left\{\begin{array}{l}
{[}e_{2},  \cdots, e_{n+1}] = e_{1}, \\
{[}e_{1}, \hat{e}_2, \hat{e}_3, \cdots, e_{n+2}] = (b^{2}_{1, 3} +b^{3}_{1, 2})e_{1}, \\
\cdots \cdots \cdots \cdots \cdots \cdots \cdots\\
{[}e_{1}, \cdots, \hat{e}_{i}, \cdots, \hat{e}_{j}, \cdots, e_{n+2}]
= (b^{i}_{1, j} +b^{j}_{1, i})e_{1},
 \\
\cdots \cdots \cdots \cdots \cdots \cdots \cdots\\
{[}e_{1}, \cdots, \hat{e}_{n}, \hat{e}_{n+1}, e_{n+2}] = (b^{n}_{1, n+1} +b^{n+1}_{1, n})e_{1}, \\
{[}\hat{e}_{2}, e_{3}, \cdots, e_{n+2}] = b^{2}_{1, 2} e_{2}+
\cdots+b^{r}_{1, 2} e_{r},\\
\cdots \cdots \cdots \cdots \cdots \cdots \cdots\\
{[}e_{2}, \cdots, \hat{e}_{k},\cdots, e_{n+2}] = b^{2}_{1, k} e_{2}+ \cdots+b^{r}_{1, k} e_{r},\\
\cdots \cdots \cdots \cdots \cdots \cdots \cdots\\
{[}e_{2}, \cdots, \hat{e}_{n+1},  e_{n+2}] = b^{2}_{1, n+1} e_{2}+ \cdots+b^{r}_{1, n+1} e_{r};\\
\end{array}\right.
\end{array}$
\vspace{2mm}\\where ~$ 2\leq i \neq j \leq n+1, ~2\leq k\leq n+1.$

 \vspace{1mm} We conclude parameters $b^{i}_{1,l}=0$ for $r< l \leq n+1, ~i=2,
3, \cdots, r.$ In fact, we might as well suppose that  there exists
$b^{2}_{1, l}\neq 0$ for some $l$ satisfying $r< l \leq n+1$, and
choose $l=min\{~k'~|~r< k' \leq n+1, ~b^{2}_{1, k'}\neq 0 \}$.
Replacing $e_{l}+\sum\limits_{m=l+1}^{n+1}\frac{b^{2}_{1,
m}}{b^{2}_{1, l}} e_{m}$ for $e_{l}$ in $(2r)'$,  we have that
$(2r)'$ can be written as

\vspace{2mm}\noindent $\begin{array}{l} \left\{\begin{array}{l}
{[}e_{2}, \cdots, e_{n+1}] = e_{1}, \\
{[}e_{1}, e_{4}, \cdots, e_{n+2}] = (b^{2}_{1, 3} +b^{3}_{1,
2})e_{1}+(\frac{b^{2}_{1, 2}(b^{2}_{1, l} +b^{l}_{1, 2})}{b^{2}_{1,
l}} +
\frac{b^{2}_{1, 3}(b^{l}_{1, 3} +b^{3}_{1, l})}{b^{2}_{1, l}})e_{1},\\
\cdots \cdots \cdots \cdots \cdots \cdots \cdots\\
{[}e_{1}, \cdots, \hat{e}_{i}, \cdots, \hat{e}_{j}, \cdots, e_{n+2}]
=
 (b^{j}_{1, i} +b^{i}_{1, j})e_{1}+(\frac{b^{2}_{1, i}(b^{i}_{1, l} +b^{l}_{1, i})}{b^{2}_{1, l}} +
 \frac{b^{2}_{1, j}(b^{l}_{1, j} +b^{j}_{1, l})}{b^{2}_{1, l}})e_{1}, \\
\cdots \cdots \cdots \cdots \cdots \cdots \cdots\\
{[}e_{1}, \cdots, \hat{e}_{n}, \hat{e}_{n+1}, e_{n+2}] =
(b^{n+1}_{1, n} +b^{n}_{1, n+1})e_{1}+(\frac{b^{2}_{1, n}(b^{n}_{1,
l} +b^{l}_{1, n})}{b^{2}_{1, l}}+
\frac{b^{2}_{1, n+1}(b^{l}_{1, n+1} +b^{n+1}_{1, l})}{b^{2}_{1, l}})e_{1}, \\
{[}\hat{e}_{2}, e_{3}, \cdots, e_{n+2}] = b^{2}_{1, 2} e_{2}+
\cdots+b^{r}_{1, 2} e_{r},\\
\cdots \cdots \cdots \cdots \cdots \cdots \cdots\\
{[}e_{2}, \cdots, \hat{e}_{k}, \cdots, e_{n+2}] = b^{2}_{1, k} e_{2}+ \cdots+b^{r}_{1, k} e_{r}, ~2\leq k\leq r,\\
\cdots \cdots \cdots \cdots \cdots \cdots \cdots\\
{[}e_{2}, \cdots, \hat{e}_{r}, \cdots, e_{n+2}] = b^{2}_{1, r} e_{2}+ \cdots+b^{r}_{1, r} e_{r},\\
{[}e_{2}, \cdots, e_r, \hat{e}_{r+1}, \cdots, e_{n+2}] = B^{3}_{1, r+1} e_{3}+ \cdots+B^{r}_{1, r+1} e_{r},\\
\cdots \cdots \cdots \cdots \cdots \cdots \cdots\\
{[}e_{2}, \cdots, e_{r}, \cdots,\hat{e}_{l-1}, \cdots, e_{n+2}] =
B^{3}_{1, l-1} e_{3}+
\cdots+B^{r}_{1, l-1} e_{r},\\
{[}e_{2}, \cdots, e_{r}, \cdots, \hat{e}_{l}, \cdots, e_{n+2}] = b^{2}_{1, l} e_{2}+ \cdots+b^{r}_{1,l} e_{r},\\
{[}e_{2}, \cdots, e_{r}, \cdots,\hat{e}_{l+1},\cdots, e_{n+2}] = B^{3}_{1, l+1} e_{3}+ \cdots+B^{r}_{1, l+1} e_{r},\\
\cdots \cdots \cdots \cdots \cdots \cdots \cdots\\
{[}e_{2}, \cdots, e_{r}, \cdots, \hat{e}_{n+1}, e_{n+2}] = B^{3}_{1,
n+1}
 e_{3}+ \cdots+B^{r}_{1, n+1} e_{r};
\end{array}\right.
\end{array} $

\noindent where $2\leq i\neq j \leq n+1$. From $[\frac{1}{b^{2}_{1,
l}}([e_{2},$ $\cdots, \hat{e}_{l},$ $\cdots, e_{n+2}]+b^{3}_{1, l}
e_{3}+\cdots+b^{r}_{1, l} e_{r}),$ $e_{3}, \cdots, $ $\hat{e}_{m},
\cdots, e_{n+2}]$ $=\frac{b^{2}_{1, m}+b^{m}_{1, 2}}{b^{2}_{1,
l}}(b^{2}_{1, l} e_{2}$ $+\cdots+b^{r}_{1, l} e_{r})+\frac{b^{m}_{1,
l}}{b^{2}_{1, l}}(b^{2}_{1, 2} e_{2}$ $+\cdots+b^{r}_{1, 2} e_{r})$
for $3\leq m\leq r$, $(2r)'$ is reduced to

\vspace{2mm}\noindent $\begin{array}{l} \left\{\begin{array}{l}
{[}e_{2},  \cdots, e_{n+1}] = e_{1}, \\
{[}e_{1}, \hat{e}_2, \hat{e}_3,  \cdots, e_{n+2}] = (b^{2}_{1, 3}
+b^{3}_{1, 2})e_{1}+(\frac{b^{2}_{1, 2}(b^{2}_{1, l} +b^{l}_{1,
2})}{b^{2}_{1, l}} +
\frac{b^{2}_{1, 3}(b^{l}_{1, 3} +b^{3}_{1, l})}{b^{2}_{1, l}})e_{1},\\
\cdots \cdots \cdots \cdots \cdots \cdots \cdots\\
{[}e_{1}, \cdots, \hat{e}_{i}, \cdots, \hat{e}_{j}, \cdots, e_{n+2}]
=
 (b^{j}_{1, i} +b^{i}_{1, j})e_{1}+(\frac{b^{2}_{1, i}(b^{i}_{1, l} +b^{l}_{1, i})}{b^{2}_{1, l}} +
 \frac{b^{2}_{1, j}(b^{l}_{1, j} +b^{j}_{1, l})}{b^{2}_{1, l}})e_{1}, \\
\cdots \cdots \cdots \cdots \cdots \cdots \cdots\\
{[}e_{1}, \cdots, \hat{e}_{n}, \hat{e}_{n+1}, e_{n+2}] =
(b^{n+1}_{1, n} +b^{n}_{1, n+1})e_{1}+(\frac{b^{2}_{1, n}(b^{n}_{1,
l} +b^{l}_{1, n})}{b^{2}_{1, l}}+
\frac{b^{2}_{1, n+1}(b^{l}_{1, n+1} +b^{n+1}_{1, l})}{b^{2}_{1, l}})e_{1}, \\
{[}e_{3}, \cdots, e_{n+2}] = b^{2}_{1, 2} e_{2}+
\cdots+b^{r}_{1, 2} e_{r},\\
 {[}e_{2}, \hat{e}_3, \cdots, e_{n+2}] =
\frac{b^{2}_{1, 3}+b^{3}_{1, 2}}{b^{2}_{1, l}}(b^{2}_{1, l} e_{2}+
\cdots+b^{r}_{1, l} e_{r})+\frac{b^{3}_{1, l}}{b^{2}_{1,
l}}(b^{2}_{1, 2} e_{2}+
\cdots+b^{r}_{1, 2} e_{r}),\\
\cdots \cdots \cdots \cdots \cdots \cdots \cdots\\
{[}e_{2}, \cdots, \hat{e}_{k}, \cdots, e_{n+2}] = \frac{b^{2}_{1,
k}+b^{k}_{1, 2}}{b^{2}_{1, l}}(b^{2}_{1, l} e_{2}+ \cdots+b^{r}_{1,
l} e_{r})+\frac{b^{k}_{1, l}}{b^{2}_{1, l}}(b^{2}_{1, 2} e_{2}+
\cdots+b^{r}_{1, 2} e_{r}),\\
\cdots \cdots \cdots \cdots \cdots \cdots \cdots\\
{[}e_{2}, \cdots, \hat{e}_{r}, \cdots, e_{n+2}] = \frac{b^{2}_{1,
r}+b^{r}_{1, 2}}{b^{2}_{1, l}}(b^{2}_{1, l} e_{2}+ \cdots+b^{r}_{1,
l} e_{r})+\frac{b^{r}_{1, l}}{b^{2}_{1, l}}(b^{2}_{1, 2} e_{2}+
\cdots+b^{r}_{1, 2} e_{r}),\\
{[}e_{2}, \cdots, e_{r}, \cdots, \hat{e}_{l}, \cdots, e_{n+2}] =
 b^{2}_{1, l} e_{2}+ \cdots+b^{r}_{1, l} e_{r};\\
\end{array}\right.
\end{array}$

\noindent where $2\leq i\neq j \leq n+1, 3\leq  k \leq r$. Then we
have $\dim A^1\leq 3$. This is a contradiction. Therefore,
$b^{2}_{1, l} = 0$ for $r<l \leq n+1$. Similarly, we have $b^{i}_{1,
l} = 0$ for $3\leq i \leq n+1$ (the proving process is omitted).
Therefore  $(2r)$  is isomorphic to

\vspace{2mm}\noindent$
\begin{array}{l}
\left\{\begin{array}{l}
{[}e_{2},  \cdots, e_{n+1}] = e_{1}, \\
{[}e_{1}, \hat{e}_2, \hat{e}_3, \cdots, e_{n+2}] = (b^{2}_{1, 3} +b^{3}_{1, 2})e_{1}, \\
\cdots \cdots \cdots \cdots \cdots \cdots \cdots\\
{[}e_{1},  \cdots, \hat{e}_{i}, \cdots, \hat{e}_{j}, \cdots, e_{n+2}] = (b^{i}_{1, j} +b^{j}_{1, i})e_{1}, \\
\cdots \cdots \cdots \cdots \cdots \cdots \cdots,\\
{[}e_{1},  \cdots, \hat{e}_{r-1}, \hat{e}_{r}, \cdots, e_{n+2}] = (b^{r-1}_{1, r} +b^{r}_{1, r-1})e_{1}, \\
{[}\hat{e}_{2}, e_{3}, \cdots, e_{n+2}] = b^{2}_{1, 2} e_{2}+
\cdots+b^{r}_{1, 2} e_{r},\\
\cdots \cdots \cdots \cdots \cdots \cdots \cdots\\
{[}e_{2},  \cdots, \hat{e}_{k}, \cdots, e_{n+2}] = b^{2}_{1, k} e_{2}+ \cdots+b^{r}_{1, k} e_{r},\\
\cdots \cdots \cdots \cdots \cdots \cdots \cdots\\
{[}e_{2},  \cdots, \hat{e}_{r}, \cdots, e_{n+2}] = b^{2}_{1, r} e_{2}+ \cdots+b^{r}_{1, r} e_{r};\\
\end{array}\right.
\end{array} $

\vspace{2mm}\noindent where $2\leq i\neq j \leq r$ and $2\leq k \leq
r.$ Since $\dim A^{1}=r$,
 for any $i, j$ satisfying $i\neq j$, $2\leq i, j\leq r$,  there exist $b^{i}_{1, s}\neq
 0$ and
$b^{j}_{1, t}\neq 0$ for some $s, t$, where $2\leq s, t \leq r,
s\neq t$. From the products

\vspace{2mm}\noindent$[\frac{1}{b^{i}_{1, s}}([e_{2},$
 $\cdots, \hat{e}_{s}, \cdots,
e_{n+2}]-\sum\limits_{k=2}^{i-1} b^{k}_{1, s}
e_{k}-\sum\limits_{k=i+1}^{r} b^{k}_{1, s} e_{k}), e_{2},$ $\cdots,$
$ \hat{e}_i, \cdots,$ $\hat{e}_{k},$ $\cdots,$ $ e_{n+2}]$,

\vspace{2mm}\noindent  and $[\frac{1}{b^{j}_{1, t}}([e_{2},$  $
\cdots,$ $ \hat{e}_{t},$ $\cdots,$ $
e_{n+2}]-\sum\limits_{k=2}^{j-1} b^{k}_{1, t}
e_{k}-\sum\limits_{k=j+1}^{r} b^{k}_{1, t} e_{k}), e_{2},$ $
\cdots,$ $\hat{e}_{j},$ $\cdots,$ $\hat{e}_{k},$ $\cdots,$ $
e_{n+2}]$, for $2\leq i\neq j \leq r$, $2\leq s\neq t \leq r$,
$2\leq k\leq r$, $k\neq s$ and $k\neq t$, we obtain $b^{s}_{1,
k}=b^{k}_{1, s}$ and $b^{t}_{1, k}=b^{k}_{1, t}$. Then the
multiplication $(2r)'$ is of the form

\vspace{2mm}\noindent $\begin{array}{l} (2r)'' ~
\left\{\begin{array}{l}
{[}e_{2},  \cdots, e_{n+1}] = e_{1}, \\
{[}\hat{e}_{2}, e_{3}, \cdots, e_{n+2}] = b^{2}_{1, 2} e_{2}+
\cdots+b^{r}_{1, 2} e_{r},\\
\cdots \cdots \cdots \cdots \cdots \cdots \cdots\\
{[}e_{2}, \cdots, \hat{e}_{k}, \cdots, e_{n+2}] = b^{2}_{1, k} e_{2}+ \cdots+b^{r}_{1, k} e_{r},\\
\cdots \cdots \cdots \cdots \cdots \cdots \cdots\\
{[}e_{2}, \cdots, \hat{e}_{r}, \cdots, e_{n+2}] = b^{2}_{1, r} e_{2}+ \cdots+b^{r}_{1, r} e_{r};\\
\end{array}\right.
\end{array}$~ $b^{s}_{1, k}=b^{k}_{1, s}$,~ $2\leq k \leq
 r$.

 Since $\dim A^1=r$, for any $k$ satisfying $2\leq k\leq r$, there exists $b^k_{1, j}\neq 0$.
 We choose $b_{1, j_2}^2\neq 0$ such that $j_2=min\{~j~|~2\leq j\leq r, ~b_{1, j}^2\neq
 0\}.$ Taking  a linear transformation of the basis $e_1, \cdots, e_{n+2}$ by replacing
$e_{j_2}+\sum\limits_{m=j_2+1}^{r}\frac{b^{2}_{1, m}}{b^{2}_{1,
j_2}} e_{m}$ for $e_{j_2}$, then $(2r)''$ can be written as

\vspace{2mm}\noindent $\begin{array}{l} (2r)^{j_2}~
\left\{\begin{array}{l}
{[}e_{2}, \cdots, e_{n+1}] = e_{1}, \\
{[}\hat{e}_{2}, e_{3}, \cdots, e_{n+2}] = C^{3}_{1, 2} e_{3}+
\cdots+C^{r}_{1, 2} e_{r},\\
\cdots \cdots \cdots \cdots \cdots \cdots \cdots\\
{[}e_{2}, \cdots, \hat{e}_{j_2-1}, \cdots, e_{n+2}] = C^{3}_{1,
j_2-1} e_{3}+
 \cdots+C^{r}_{1, j_2-1} e_{r},\\
{[}e_{2}, \cdots, \hat{e}_{j_2}, \cdots, e_{n+2}] = C^{2}_{1, j_2} e_{2}+ \cdots+C^{r}_{1, j_2} e_{r},\\
{[}e_{2}, \cdots, \hat{e}_{j_2+1}, \cdots, e_{n+2}] = C^{3}_{1, j_2+1} e_{3}+ \cdots+C^{r}_{1, j_2+1} e_{r},\\
\cdots \cdots \cdots \cdots \cdots \cdots \cdots\\
{[}e_{2}, \cdots, \hat{e}_{r}, \cdots, e_{n+2}] = C^{3}_{1, r} e_{3}+ \cdots+C^{r}_{1, r} e_{r}.\\
\end{array}\right.
\end{array}$

By introduction, replacing $e_{j_l}$ by
$e_{j_l}+\sum\limits_{m=j_l+1}^{r}\frac{C^{l}_{1, m}}{b^{l}_{1,
j_l}} e_{m}$ in  $(2r)^{j_{l-1}}$ for $3\leq l\leq r$, where $C_{1,
j_l}^l\neq 0$ and
 $j_l=min\{~j~|~2\leq j\leq r, ~C_{1, j}^l\neq
 0, ~j\neq j_2, \cdots, j_{l-1}\}$,  we get the reduced form of
 $(2r)''$ as follows

\vspace{2mm}\noindent $\begin{array}{l} ~ \left\{\begin{array}{l}
{[}e_{2}, \cdots, e_{n+1}] = e_{1}, \\
{[}e_{2}, \cdots, \hat{e}_{j_{2}}, \cdots, e_{n+2}] = D^{2}_{1,
j_{2}} e_{2} +
\cdots + D^{r}_{1, j_{2}} e_{r},\\
{[}e_{2}, \cdots, \hat{e}_{j_{3}}, \cdots, e_{n+2}] = D^{3}_{1,
j_{3}} e_{3} +
\cdots + D^{r}_{1, j_3} e_{r},\\
\cdots \cdots \cdots \cdots \cdots \cdots \cdots\\
{[}e_{2}, \cdots, \hat{e}_{j_{r-1}}, \cdots, e_{n+2}] =
D^{r-1}_{1, j_{r-1}} e_{r-1}+D^{r}_{1, j_{r-1}} e_{r},\\
{[}e_{2}, \cdots, \hat{e}_{j_{r}}, \cdots, e_{n+2}] = D^{r}_{1,
j_{r}} e_{r};
\end{array}\right.
\end{array} $

\vspace{1mm}\noindent where  $\{j_{2}, j_{3}, \cdots, j_{r}\}$ $=
\{2, 3, \cdots, r\}$ and $D^{k}_{1, j_{k} }$ $\neq 0$ for $ 2\leq
k\leq r.$

From  the products $[\frac{1}{D^{k}_{1, j_{k}}}([e_{2},$
 $\cdots,$ $\hat{e}_{j_{k}},$ $\cdots,$
$e_{n+2}]+D^{k+1}_{1, j_{k}} e_{k+1}+\cdots +D^{r}_{1, j_{k}}
e_{r}), e_{2},$  $ \cdots,$ $\hat{e}_{j_{i}},$ $\cdots,$
$\hat{e}_{k},$ $\cdots,$ $ e_{n+2}]$ for $k\leq r$,  we get
$D^{j_{k}}_{1, j_{i}}=D^{j_{i}}_{1, j_{k}}=0$ when $j_{i}, j_{k}$
satisfy conditions $j_{i}\neq j_{k}$, $j_{i}\neq k$ and $j_{k}\neq
i$. Since $r$ is even,  $(2r)'$ is of the form

\vspace{2mm}\noindent $\begin{array}{l} (2r)'''~
\left\{\begin{array}{l}
{[}e_{2},  \cdots, e_{n+1}] = e_{1}, \\
{[}\hat{e}_{2}, e_{3}, \cdots, e_{n+2}] = D^{2}_{1, 2} e_{2}, \\
\cdots \cdots \cdots \cdots \cdots \cdots \cdots\\
{[}e_{2}, \cdots, \hat{e}_{i}, \cdots, e_{n+2}] = D^{i}_{1, i} e_{i}, ~~D^{i}_{1, i}\neq0, ~2\leq i\leq r,\\
\cdots \cdots \cdots \cdots \cdots \cdots \cdots\\
 {[}e_{2}, \cdots,\hat{e}_{r},\cdots, e_{n+2}]
= D^{r}_{1, r}e_{r}.
\end{array}\right.
\end{array}
 $

\vspace{1mm}Replacing $e_{n+2}$ and $e_r$ by $\frac{1}{D^{r}_{1,
r}}e_{n+2}$ and $\frac{D^{r}_{1, r}}{D^{r-1}_{1, r-1}}e_{r}$ in
$(2r)'''$, we get

\vspace{2mm}\noindent $
\begin{array}{l}
\left\{\begin{array}{l}
{[} e_{2}, \cdots, e_{n+1}] = e_{1}, \\
{[}\hat{e}_{2}, e_{3}, \cdots, e_{n+2}] = \alpha_{2} e_{2}, \\
\cdots \cdots \cdots \cdots \cdots \cdots \cdots\\
{[}e_{3},\cdots, \hat{e}_{i}, \cdots, e_{n+2}] = \alpha_{i} e_{i}, \\
\cdots \cdots \cdots \cdots \cdots \cdots \cdots\\
{[}e_{2}, \cdots, \hat{e}_{r-2}, \cdots, e_{n+2}] = \alpha_{r-2}e_{r-2}, \\
{[}e_{2}, \cdots, \hat{e}_{r-1}, \cdots, e_{n+2}] = e_{r-1}, \\
{[}e_{2}, \cdots, \hat{e}_{r}, \cdots, e_{n+2}] = e_{r};
\end{array}\right.
\end{array} ~ \alpha_{i}\in F,
 ~\alpha_{i}\neq 0, ~2\leq i\leq r-2.$

\vspace{1mm} Now let $B=A$ as  vector spaces. Defining an
$(r-2)$-ary multiplication $[, \cdots, ]_1$ on $B$ by fixing
$e_{r+1}, \cdots, e_{n+2}$ in the multiplication of $A$, that is for
$x_1, \cdots, x_{r-2}\in B,$ $[x_1, \cdots, x_{r-2}]_1=[x_1, \cdots,
x_{r-1}, e_{r+1}, \cdots, e_{n+2}]$. Then $B$ is an $(r-2)$-Lie
algebra and has the decomposition ~$ B=W\oplus Z(B),$~ where
$W=Fe_2+\cdots+Fe_{r}$  is an $(r-1)$-dimensional ideal of $B$ with
$W=W^1=Fe_2+\cdots+Fe_{r}=B^1$, and
$~Z(B)=Fe_1+Fe_{r+1}+\cdots+Fe_{n+2} $ is the center of $B$. Thanks
to Lemma 2.1, there exists a basis $e'_2, \cdots, e'_{r}$ of $W$
such that the multiplication table of $W$ is $(d_1)$ or $(d_2)$ of
Lemma 2.1. Therefore, we can choose a suitable basis $e_1, \cdots
e_{n+2}$ and write the table (2r) as

\vspace{2mm}\noindent$
\begin{array}{l}
(e^{3})~  \left\{\begin{array}{l}
{[} e_{2}, \cdots,e_{n+1}] = e_{1}, \\
{[}\hat{e}_{2}, e_{3},  \cdots, e_{n+2}] = e_{2}, \\
\cdots \cdots \cdots \cdots \cdots \cdots \cdots\\
{[}e_{2}, \cdots, \hat{e}_{i},  \cdots, e_{n+2}] = e_{i},  \\
\cdots \cdots \cdots \cdots \cdots \cdots \cdots\\
{[}e_{2}, \cdots, \hat{e}_{r}, \cdots, e_{n+2}] = e_{r};
\end{array}\right.
\end{array}~ 2\leq i\leq r.
$

It is not difficult to see that  $(e^{1})$ and $(e^2)$ are not
isomorphic to $(e^{3})$  since $(e^{1})$ and $(e^{2})$ are
decomposable. And by Lemma 2.1, $(e^{1})$ is not isomorphic to
$(e^{2})$.

$\bar{5}$ ~  If $\dim A^{1}=r\geq 5$ and  $r$ is odd. Suppose $A^1=F
e_1+F e_2+\cdots+F e_r$. Then the multiplication table of $A$ in the
basis $e_1, \cdots, e_{n+2}$ has only following possibilities

\vspace{2mm}\noindent $\begin{array}{l} (1)~ \left\{\begin{array}{l}
{[}\hat{e}_{1}, e_{2}, \cdots,e_{n+1}] = e_{1}, \\
\cdots \cdots \cdots \cdots \cdots \cdots \cdots\\
{[}e_{1}, \cdots, \hat{e}_{p}, \cdots, e_{n+1}] = e_{p}, \\
{[}e_{1}, \cdots, \hat{e}_{p+1}, \cdots, e_{n+1}] = e_{r}, \\
{[}e_{1}, \cdots, \hat{e}_{p+2}, \cdots, e_{n+1}] = e_{r-1}, \\
\cdots \cdots \cdots \cdots \cdots \cdots \cdots\\
{[}e_{1}, \cdots, \hat{e}_{p+q}, \cdots, e_{n+1}] = e_{p+1},\\
{[}e_{1}, \cdots, \hat{e}_{i}, \cdots, \hat{e}_{j}, \cdots, e_{n+2}]
=\sum\limits_{k=1}^{r}b^{k}_{i, j} e_{k};
\end{array}\right.
\end{array}  p+q=r,  ~2\leq  q < r; $

\vspace{2mm}\noindent$\begin{array}{l} (2)~ \left\{\begin{array}{l}
{[} \hat{e}_1, e_{2}, \cdots, e_{n+1}] = e_{1}, \\
\cdots \cdots \cdots \cdots \cdots \cdots \cdots\\
{[}e_{1}, \cdots, \hat{e}_{k}, \cdots, e_{n+1}] = e_{k}, \\
\cdots \cdots \cdots \cdots \cdots \cdots \cdots\\
{[}e_{1}, \cdots, \hat{e}_{r}, \cdots, e_{n+1}] = e_{r},\\
{[}e_{1}, \cdots, \hat{e}_{i}, \cdots, \hat{e}_{j}, \cdots, e_{n+2}]
=\sum\limits_{k=1}^{r}b^{k}_{i, j} e_{k};
\end{array}\right.
\end{array}~1\leq k\leq r;$

\vspace{2mm}\noindent$\begin{array}{l} (3)~ \left\{\begin{array}{l}
{[}\hat{e}_{1}, e_{2}, \cdots,e_{n+1}] = e_{1}, \\
\cdots \cdots \cdots \cdots \cdots \cdots \cdots\\
{[}e_{1}, \cdots, \hat{e}_{p}, \cdots, e_{n+1}] = e_{p}, \\
{[}e_{1}, \cdots, \hat{e}_{p+1}, \cdots, e_{n+1}] = e_{r-1}, \\
{[}e_{1}, \cdots, \hat{e}_{p+2}, \cdots, e_{n+1}] = e_{r-2}, \\
\cdots \cdots \cdots \cdots \cdots \cdots \cdots\\
{[}e_{1}, \cdots, \hat{e}_{p+q}, \cdots, e_{n+1}] = e_{p+1}, \\
{[}e_{1}, \cdots, \hat{e}_{i}, \cdots, \hat{e}_{j}, \cdots, e_{n+2}]
=\sum\limits_{k=1}^{r}b^{k}_{i,j} e_{k};
\end{array}\right.
\end{array} 2\leq q < r, ~ p+q=r-1;$

\vspace{2mm}\noindent$\begin{array}{l} (4)~  \left\{\begin{array}{l}
{[}\hat{e}_1, e_{2}, \cdots, e_{n+1}] = e_{1}, \\
\cdots \cdots \cdots \cdots \cdots \cdots \cdots\\
{[} e_{1},  \cdots, \hat{e}_k, \cdots, e_{n+1}] = e_{k},  \\
\cdots \cdots \cdots \cdots \cdots \cdots \cdots\\
{[}e_{1}, \cdots, \hat{e}_{r-1}, \cdots, e_{n+1}] = e_{r-1},\\
{[}e_{1}, \cdots, \hat{e}_{i}, \cdots, \hat{e}_{j}, \cdots, e_{n+2}]
=\sum\limits_{k=1}^{r}b^{k}_{i, j} e_{k};
\end{array}\right.
\end{array}~1\leq k\leq r-1;$

\vspace{2mm}\noindent$\cdots \cdots \cdots \cdots \cdots \cdots
\cdots\cdots \cdots \cdots \cdots \cdots \cdots \cdots$

\vspace{2mm}\noindent$\begin{array}{l} (2r-3)~
\left\{\begin{array}{l}
{[}e_{2}, \cdots, e_{n+1}] = e_{1}, \\
{[}e_{1}, e_{3}, \cdots, e_{n+1}] = e_{2}, \\
{[}e_{1}, \cdots, \hat{e}_{i}, \cdots, \hat{e}_{j}, \cdots, e_{n+2}]
=\sum\limits_{k=1}^{r}b^{k}_{i, j} e_{k};
\end{array}\right.
\end{array} $

\vspace{2mm}\noindent$\begin{array}{l} (2r-2)~
\left\{\begin{array}{l}
{[}e_{2}, \cdots, e_{n+1}] = e_{1}+\alpha e_{2},\\
{[}e_{1}, e_{3}, \cdots, e_{n+1}] = e_{2}, \\
{[}e_{1}, \cdots, \hat{e}_{i}, \cdots, \hat{e}_{j}, \cdots, e_{n+2}]
=\sum\limits_{k=1}^{r}b^{k}_{i, j} e_{k};
\end{array}\right.
\end{array} $

\vspace{2mm}\noindent $\begin{array}{l} (2r-1)~
\left\{\begin{array}{l}
{[}e_{1}, \cdots, e_{n}] = e_{1},\\
{[}e_{1},  \cdots, \hat{e}_{i},\cdots, \hat{e}_{j}, \cdots, e_{n+2}]
=\sum\limits_{k=1}^{r}b^{k}_{i, j} e_{k};
\end{array}\right.
\end{array} $

\vspace{2mm}\noindent $\begin{array}{l} (2r)~~
\left\{\begin{array}{l}
{[}e_{2},  \cdots, e_{n+1}] = e_{1},\\
{[}e_{1}, \cdots, \hat{e}_{i}, \cdots, \hat{e}_{j}, \cdots, e_{n+2}]
=\sum\limits_{k=1}^{r}b^{k}_{i, j} e_{k};
\end{array}\right.
\end{array} $

\vspace{2mm}\noindent $\begin{array}{l} (2r+1)~
\left\{\begin{array}{l} {[}\hat{e}_1, \hat{e}_2,  \cdots, e_{r+1},
\cdots, e_{n+2}]
=b^{1}_{1, 2} e_{1}+b^{2}_{1, 2} e_{2}+\cdots+b^{r}_{1, 2} e_{r},\\
\cdots \cdots \cdots \cdots \cdots \cdots \cdots\\
{[}e_{1}, \cdots, \hat{e}_{k}, \cdots, \hat{e}_{l}, \cdots, e_{r+1},
\cdots, e_{n+2}]
=b^{1}_{k, l} e_{1}+b^{2}_{k, l} e_{2}+\cdots+b^{r}_{k, l} e_{r},\\
\cdots \cdots \cdots \cdots \cdots \cdots \cdots\\
{[}e_{1}, \cdots, \hat{e}_{r-1}, \hat{e}_{r},  e_{r+1}, \cdots,
e_{n+2}] =b^{1}_{r-1, r} e_{1}+b^{2}_{r-1, r}
e_{2}+\cdots+b^{r}_{r-1, r} e_{r}.
\end{array}\right.
\end{array}$

\vspace{2mm}\noindent where ~$q$ is even,  ~$ i\neq j, 1\leq i,
j\leq n+1, $ ~$ k\neq l, ~1\leq k, ~l\leq r.$

Similar  to the case when $r$ being even, the cases $(3), \cdots
(2r-1)$ and the case (2r+1) are not realized. From the table (1) and
(2) we obtain the  non-isomorphic classes $(\bar{e}^{1})$,
$(\bar{e}^{2})$, $(\bar{e}^{3})$ and $(\bar{e}^{4})$. And  from the
table $(2r)$ we obtain the non-isomorphic classes $(\bar{e}^{5})$
and $(\bar{e}^{6})$. And $(\bar{e}^i)$ is not isomorphic to
$(\bar{e}^j)$ for $i=1, 2, 3, 4$, $j=5, 6$. \hfill$\Box$.

\vspace{2mm}\noindent {\bf Acknowledgments:} The first author would
like to thank  School of Mathematics and Physics at The University
of Queensland for warm hospitality. The third author acknowledges
the support of the Australian Research Council.

\vspace{5mm}\noindent {\bf References}

\begin{description}

\item{[1]} V. Filippov, $n$-Lie algebras,  Sib. Mat. Zh., 1985, 26 (6), 126-140.

\item{[2]} W. Ling. On the structure of $n$-Lie
                  algebras, Dissertation, University-GHS-Siegen, Siegn 1993.
\item{[3]}  A. Pozhidaev, Simple quotient algebras and
             subalgebras of Jacobian algebras, Sib. Math. J., 1998, 39(3), 512-517.
\item{[4]}  A. Pozhidaev,  Monomial n-Lie algebras, Algebra
             i Logika, 1998, 37(5), 307-322.
\item{[5]}  A. Pozhidaev,  On simple n-Lie algebras, Algebra
             i Logika, 1999, 38(3), 181-192.
\item{[6]} R. Bai, X. Wang, W. Xiao and H. An. The structure of low dimensional
                $n$-Lie  algebras over a field of characteristic $2$.
                Linear Algebra and Its Application, 2008, 428: 1912-1920.
\item{[7]} D.W. Barnes, On $(n+2)$ dimensional $n$-Lie algebras, arXiv: 0704.1892.
\item{[8]} W. A. de Graaf,  Classification of solvable Lie algebras,
               Experimental Mathematics, 2005, 14: 15-25.
\item{[9]} James E Humphreys,  Introduction to Lie algebras and representation theory,
                New York: Springer-Verlag, 1972.

\end{description}
\end{document}